\documentclass[review]{elsarticle}
\usepackage{tabularx}
\usepackage{lineno,hyperref}
\usepackage{adjustbox,lipsum}
\modulolinenumbers[5]
\journal{Journal of Advances in High Energy Physics}

\begin{document}
\begin{frontmatter}
\title{A new distribution for multiplicities in leptonic and hadronic collisions at high energies}
%\tnotetext[mytitlenote]{Fully documented templates are available in the elsarticle 
%package on \href{http://www.ctan.org/tex-archive/macros/latex/contrib/elsarticle}{CTAN}.}
\author{R.Chawla}
\ead{Ridhi.Chawla1988@gmail.com}
\author{M.Kaur\fnref{myfootnote}}
\ead{manjit@pu.ac.in}
\address{Physics Department, Panjab University, Chandigarh, India-160014}

%author[mymainaddress,mysecondaryaddress]{Elsevier Inc}
%ead[url]{www.elsevier.com}
%author[mysecondaryaddress]{Global Customer Service\corref{mycorrespondingauthor}}
%cortext[mycorrespondingauthor]{M.Kaur}
%ead{manjit@pu.ac.in}

%address[mymainaddress]{1600 John F Kennedy Boulevard, Philadelphia}
%address[mysecondaryaddress]{360 Park Avenue South, New York}

\begin{abstract}
Charged particles production in the $e^{+}e^{-}$, $\overline{p}p$ and $pp$ collisions in full phase space as well as in the restricted phase space slices, at high energies are described with predictions from shifted Gompertz distribution, a model of adoption of innovations.~The distribution has been extensively used in diffusion theory, social networks and forecasting.~A two-component model in which PDF is obtained from the superposition of two shifted Gompertz distributions has been introduced to improve the fitting of the experimental distributions by several orders.~The two-components correspond to the two subgroups of a data set, one representing the soft interactions and the other semi-hard interactions.~Mixing is done by appropriately assigning weights to each subgroup.~Our first attempt to analyse the data with shifted Gompertz distribution has produced extremely good results.~It is suggested that the distribution may be included in the host of distributions more often used for the multiplicity analyses.
\end{abstract}

\begin{keyword}
%\texttt{elsarticle.cls}\sep \LaTeX\sep Multiplicities \sep template
Charged multiplicities, Probability Distribution Functions, scaling violation
%\MSC[2010] 00-01\sep  99-00
\end{keyword}

\end{frontmatter}

%\linenumbers

\section{Introduction}
The shifted Gompertz distribution was introduced by Bemmaor \cite {Bema} in 1994 as a model of adoption of innovations.~It is the distribution of the largest of two independent random variables one of which has an exponential distribution with parameter $b$ and the other has a Gumbel distribution, also known as log-Weibull distribtion, with parameters $\eta$ and $b$.~Several of its statistical properties have been studied by Jim\'{e}nez and Jodr\'{a} \cite{Jod} and Jim\'{e}nez Torres \cite{Jim}.~In machine learning, the Gumbel distribution is also used to generate samples from the generalised Benoulli distribution, which is a discrete probability distribution that describes the possible results of a random variable that can take on one of the K-possible elementary events, with the probability of each elementary event separately specified.~The shifted Gompertz distribution has mostly been used in the market research and diffusion theory, social networks and forecasting.~It has also been used to predict the growth and decline of social networks, on-line services and shown to be superior to the Bass model and Weibull distribution \cite{Bau}.~It is interesting to study the statistical phenomena in high energy physics in terms of this distribution.~Recently, Weibull distribution has been used to understand the multiplicity distributions in various particle-particle collisions at high energies and more recently \cite{Wei} to explain LHC data.
Weibull models studied in the literature were appropriate for modelling a continuous random variable which assumes that the variable takes on real values over the interval [0, $\infty$].~In situations where the observed data values are very large, a continuous distribution is considered an adequate model for the discrete random variable, for example in case of a particle collider, the luminosity during a fill decreases roughly exponentially. Therefore, the mean collision rate will likewise decrease.~That decrease  will be reflected in the number of observed particles per unit time.~In the same way a photon detector that counts photons in a continuous train of time bins.~If the photons are anti-bunched in time, that is, they tend to be separated from each other, one will get a different distribution of photon counts than if the photons are bunched, that is, bunched together in time. By analyzing the photon counting statistics one can infer information about the continuous underlying distribution of the temporal spacing of photons.~The shifted Gompertz distribution with non-negative fit parameters identified with the scale and shape parameters, can in this way be used for studying the distributions of particles produced in collisions at accelerators.
One of the studies in statistics is when the variables take on discrete values. The idea was first introduced by Nakagawa and Osaki \cite{Naka}, as they introduced discrete Weibull distribution with two shape parameters $q$ and $\beta$ where $0 < q < 1 $ and $\beta > 0$.~Models which assume only non-negative integer values for modelling discrete random variables, are useful for modelling the kind of problems mentioned above.

The charged-particle multiplicity is one of the simplest observables in collisions of high energy particles, yet it imposes important constraints on the dynamics of particle production.~The particle production has been studied in terms of several theoretical, phenomenological and statistical models.~Each of these models has been reasonably successful in explaining the results from different experiments and useful for extrapolations to make predictions.~Although Weibull distribution has been studied recently, no attempt has been made so far to analyze the high energy collision data in terms of shifted Gompertz distribution.~Our first attempt to analyse the data produced good results and encouraged us for a comprehensive analysis. 

The aim of the present work is to introduce a statistical distribution, the shifted Gompertz distribution to investigate the multiplicity distributions of charged particles produced in $e^{+}e^{-}$, $pp$ and $\overline{p}p$ collisions at different center of mass energies in full phase space as well as in  restricted phase space windows.~Energy-momentum conservation strongly influences the multiplicity distribution for the full phase space.~The distribution in restricted rapidity window however, is less prone to such constraints and thus can be expected to be a more sensitive probe to the underlying dynamics of QCD, as inferred in references \cite{Urmi, Hegi}.

In Section II, details of Probability Distribution Function (PDF) of the shifted Gompertz distribution is discussed.~For $e^{+}e^{-}$ collisions a two component model has been used and modification of distributions done in terms of these two components; one from soft events and another from semi-hard events.~Superposition of distributions from these two components, by using appropriate weights is done to build the full multiplicity distribution.~When multiplicity distrbution is fitted with the weighted superposition of two shifted Gompertz distributions, we find that the agreement between data and the model improves considerably.~The fraction of soft events, $\alpha$ for various energies have been taken from references \cite{KT, DURHAM} which use the $K_{T}$ clustering algorithm, the most extensively used algorithm for LEP $e^{+}e^{-}$ data analyses.~The corresponding fractions for $pp$ and $\overline{p}p$ are not available in different rapidity bins.~For $\overline{p}p$ data at all energies under study, the $\alpha$ values for full phse space are taken from reference \cite{NBD}.~We also tried to fit the multiplicity distribution to find the best fit alpha value.~It is found that, the fit values agree very closely with values obtained from reference \cite{NBD}.~We thus fitted distributions in restricted rapidity windows for $\overline{p}p$ and $pp$ data in terms of soft and semi-hard components to get the best fit $\alpha$ values.

In a recent publication, Wilk and W\'{l}odarczyk \cite{Wilk} have developed a method of retrieving additional information from the multiplicity distributions.~They propose, in case of a conventional Negative Binomial Distribution fit \cite{NBD}, to make the parameters dependent on the multiplicity in place of having a 2-component model.~They demonstrated that the additional valuable information from the MDs, namely the oscillatory behaviour of the counting statistics can be derived.~In a future extension of the present work, we shall analyse the shifted Gompertz distribution, using the approach proposed and described by the authors \cite{Wilk}.  

~Section III presents the analysis of experimental data and the results obtained by the two approaches.~Discussion and conclusion are presented in Section IV. 

\section{Shifted Gompertz distribution}

The dynamics of hadron production can be probed using the charged particle multiplicity distribution.~Measurements of multiplicity distributions provide relevant constraints for particle-production models.~Charged particle multiplicity is defined as the average number of charged particles, $n$ produced in a collision $<n>=\sum\limits_{n=0}^{n_{max}} {nP_n}$.~Hadron production depends upon the center of mass energy available for particle production nearly independent of the types of particles undergoing collisions.~Subsequently, it is the fragmentation of quarks and gluons which produce hadrons non-perturbatively.~Thus the same PDFs can be used to describe behaviour of multiplicity distributions.~In numerous works in the past, the most popular Negative Binomial Distribution has been successfully used for a wide variety of collisions \cite{Carru}.~Universality of multiparticle production in $e^+e^-$, $pp$ and $\overline{p}p$ have been discussed in several papers, a detailed paper amongst these is \cite{Diso}.

~We briefly outline the probability density function~(PDF) of the shifted Gompertz distribution used for studying the multiplicity distributions.~Equation~(1-3) define the PDF and the mean value of the distribution;

\begin{equation}
P(n|b,\eta) = b e^{-bn}e^{-\eta e^{-bn}}[1+\eta(1-e^{-bn}]\hspace{0.3cm} for\hspace{0.15cm} n > 0 \, \label{one}
\end{equation}  

Mean of the distribution is given by
\begin{equation}
 (-\frac{1}{b})(E[ln(X)]-ln(\eta)) \hspace*{0.3 cm} where \hspace*{0.2 cm} X = \eta e^{-bn}
\end{equation}
and 
\small
\begin{equation}
 E[ln(X)] = [1 + \frac{1}{\eta}]\int_{0}^{\infty}e^{-X}[ln(X)]dX 
             - \frac{1}{\eta}\int_{0}^{\infty}Xe^{-X}[ln(X)]dX 
\end{equation}
Where $b\ge0$ is a scale parameter and $\eta \ge 0$ is a shape parameter.~Similar to the Weibull distribution, shifted Gompertz distributions is also a two parameter distribution, in terms of its shape and scale.

\subsection{ Two-Component Approach}
It is well established that at high energies, charged particle multiplicity distribution in full phase space becomes broader than a Poisson distribution.~This behaviour has been successfully described by a two parameters negative binomial (NB) distribution defined by;
\small
\begin{equation}
P(n|<n>,k)=\frac{\Gamma(n+k)}{\Gamma(n+1)\Gamma(k)}\frac{(<n>/k)^n}{(1+<n>/k)^{n+k}} 
\end{equation}
\normalsize
where $k$ is related to the dispersion $D$ by
\small
\begin{equation}
\frac{D^2}{<n>^2}=\frac{1}{<n>}+\frac{1}{k}
\end{equation} 
\normalsize
$k$ parameter of the distribution is negative in the lower energy domain where the distribution is binomial like.~$k$ is positive in the higher energy domain and the distribution is truly NB, the two particle correlations dominate and $1/k$ is closely related to the integral over full phase space of the two particle correlation function.~NB distribution was very successful until the results from UA5 collaboration \cite{UA51} showed a shoulder structure in the multiplicity distribution on $\overline{p}p$ collisions.~To expain this NB regularity violations, C. Fuglesang \cite{Fu}, suggested the violations as the effect of the weighted superposition of soft events (events without mini-jets) and semi-hard events (events with mini-jets), the weight $\alpha$ being the fraction of soft events.~The multiplicity distribution of each component being NB.~This idea was successfully implemented in several analyses at high energies to fit the multiplicity distributions with superposed NB functions.
 
Adopting this suggestion for the multiplicity distributions in $e^+e^-$, $pp$ and $\overline{p}p$ collisions at high energies, we have used a superposition of two shifted Gompertz components.~The two components are interpreted as soft and hard components, as explained above.~The Multiplicity distribution is produced by adding weighted superposition of multiplicity in soft events and multiplicity distribution in semi-hard events.~This approach combines two classes of events, not two different particle-production mechanisms in the same event.~Therefore, no interference terms are needed to be introduced.~The final distribution is the sum of the two independent distributions, henceforth called modifed shifted Gompertz distribuion.
\small
\begin{equation}
P(n)=\alpha P_{soft}^{shGomp}(n)+(1-\alpha)P_{semi-hard}^{shGomp}(n)
\end{equation} 
\normalsize
In this approach, the multiplicity distribution depends on five parameters as given below;
\small
\begin{equation}
P_{n}(\alpha:b_1,\eta_1;b_2,\eta_2)=
\alpha P_{n}(soft) + 
(1-\alpha)P_{n}(semi\textrm{-}hard) 
\end{equation} 
\normalsize
As described by A. Giovannini et al \cite{NBD}, that the superimposed physical substructures in the cases of $e^{+}e^{-}$ annihilation and hadron-hadron interactions, are different, the weighted superposition mechanism is the same.

\section{The Data}
The data from different experiments and three collision types are considered;\\
i) $e^{+}e^{-}$ annihilations at different collision energies, from 91 GeV up to the highest energy of 206.2 GeV at LEP2, from two experiments L3 \cite{L3} and OPAL \cite{OPAL91,OPAL1,OPAL2,OPAL3} are analysed.\\
ii) $pp$ collisions at LHC energies from 900 GeV, 2360 GeV and 7000 GeV \cite{CMS} are analysed in five restricted rapidity windows, $|y|$= 0.5,~1.0,~1.5,~2.0~and~2.4.\\
iii) $\overline{p}p$ collisions at energies from  200 GeV, 540 GeV and 900 GeV \cite{UA51,UA52} are analysed in full phase space as well as in restricted rapidity windows, $|y|$= 0.5,~1.5,~3.0~and~5.0.
\subsection{Results and discussion}
The PDF defined by equations (1,6) are used to fit the experimental data.~Figures~1-2 ~show the shifted Gompertz function and the modified (two-component)shifted Gompertz function fits to the data $e^{+}e^{-}$ from L3 and OPAL experiments.~Parameters of the fits, $\chi^{2}/ndf$ and the p-values are documented in Table~I.~Figure~3 shows the ratio of data over modified shifted Gompertz fit plots for $e^{+}e^{-}$ collisions at two energies.~The plots correspond to the worst and the best fits depending upon the maximum and minimum $\chi^{2}/ndf$ values and show that fluctuations between the data and the fits are acceptably small, as the ratio is nearly one.

\begin{figure}[ht]
\includegraphics[width=4.8 in, height =2.8 in]{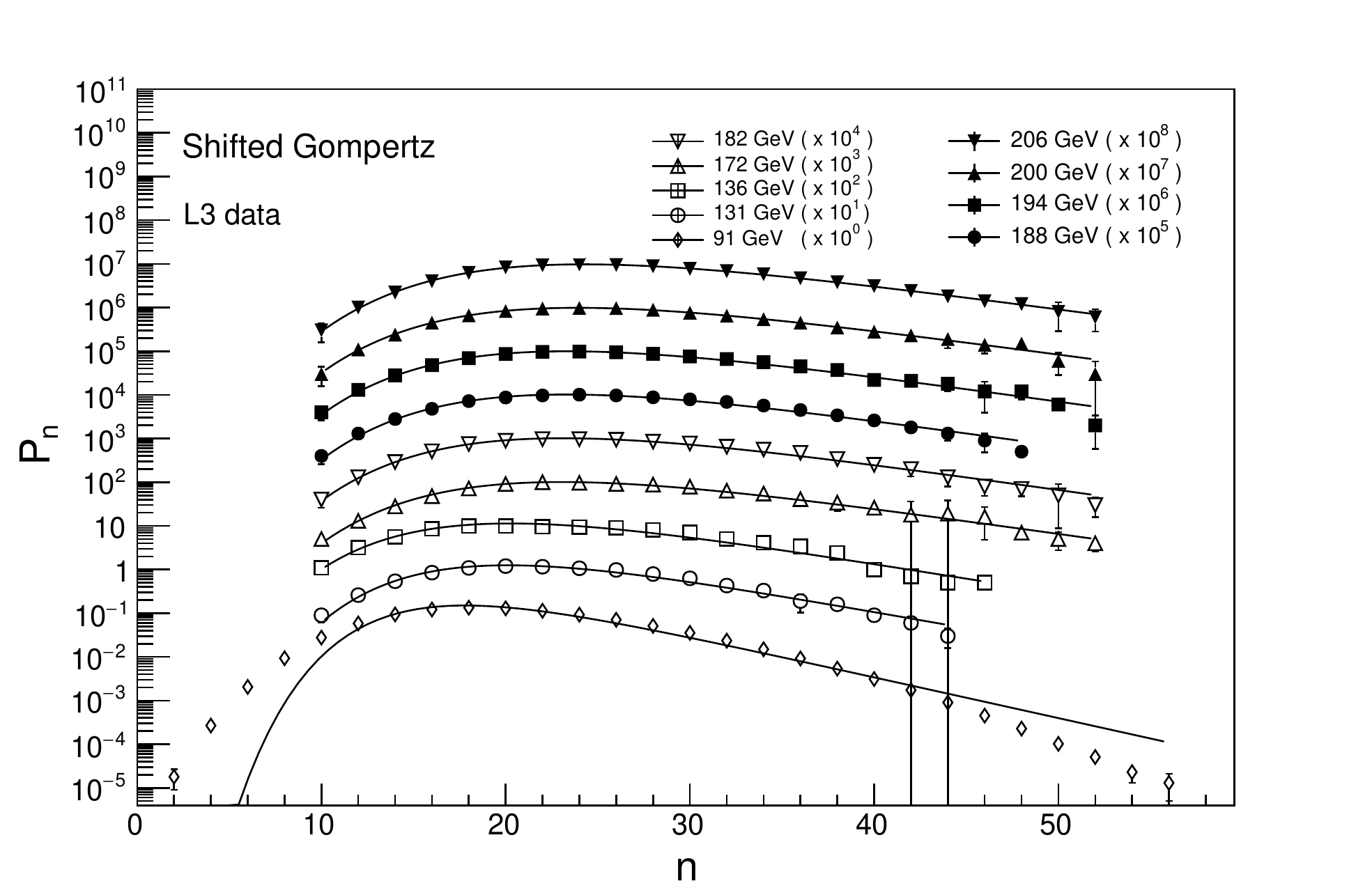}
\includegraphics[width=4.8 in, height =2.8 in]{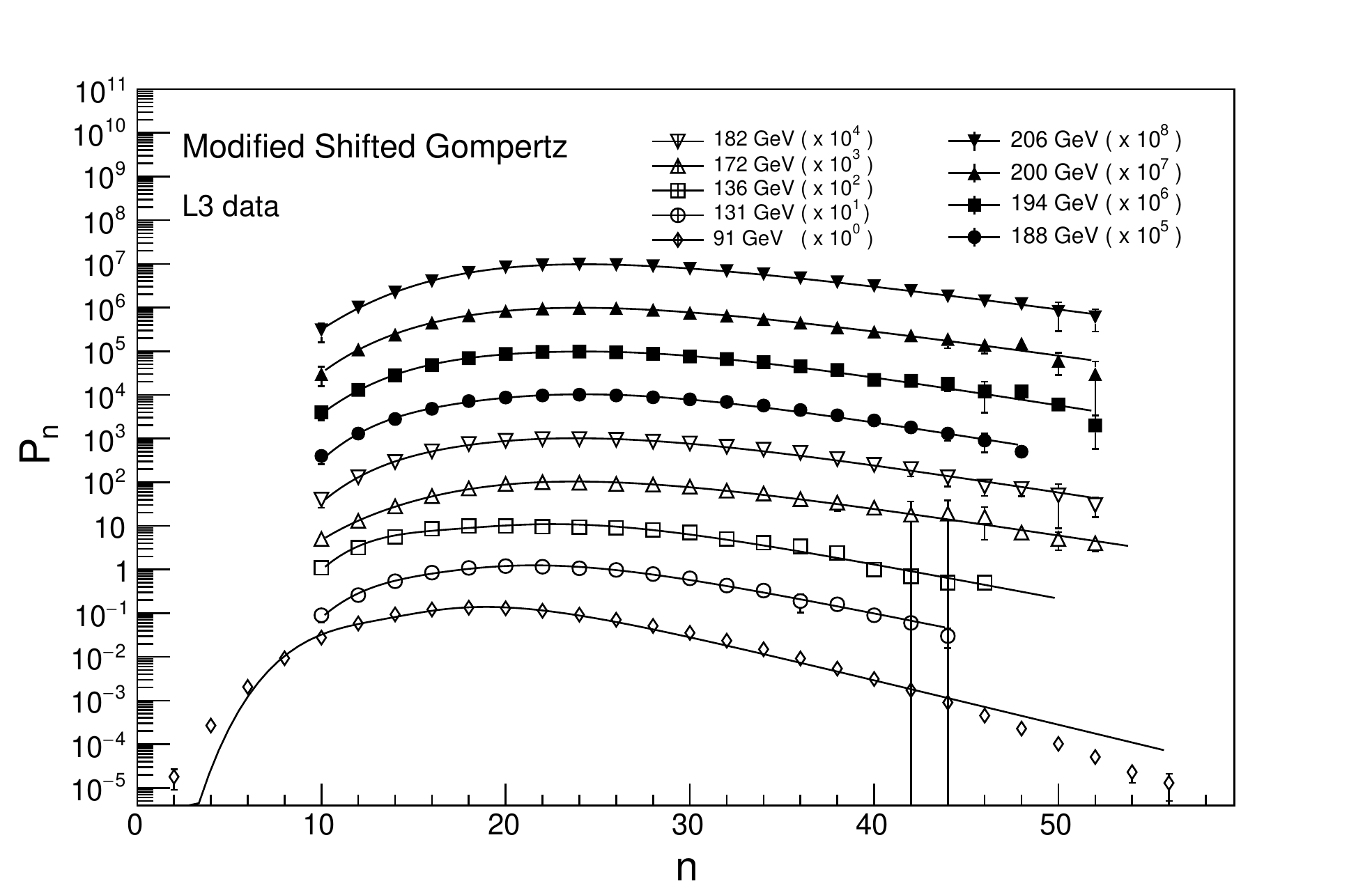}
\caption{Charged multiplicity distribution from L3 experiment.~Solid lines represent the Gompertz distribution.}
\end{figure}

\begin{figure}[ht]
\includegraphics[width=4.8 in, height =2.8 in]{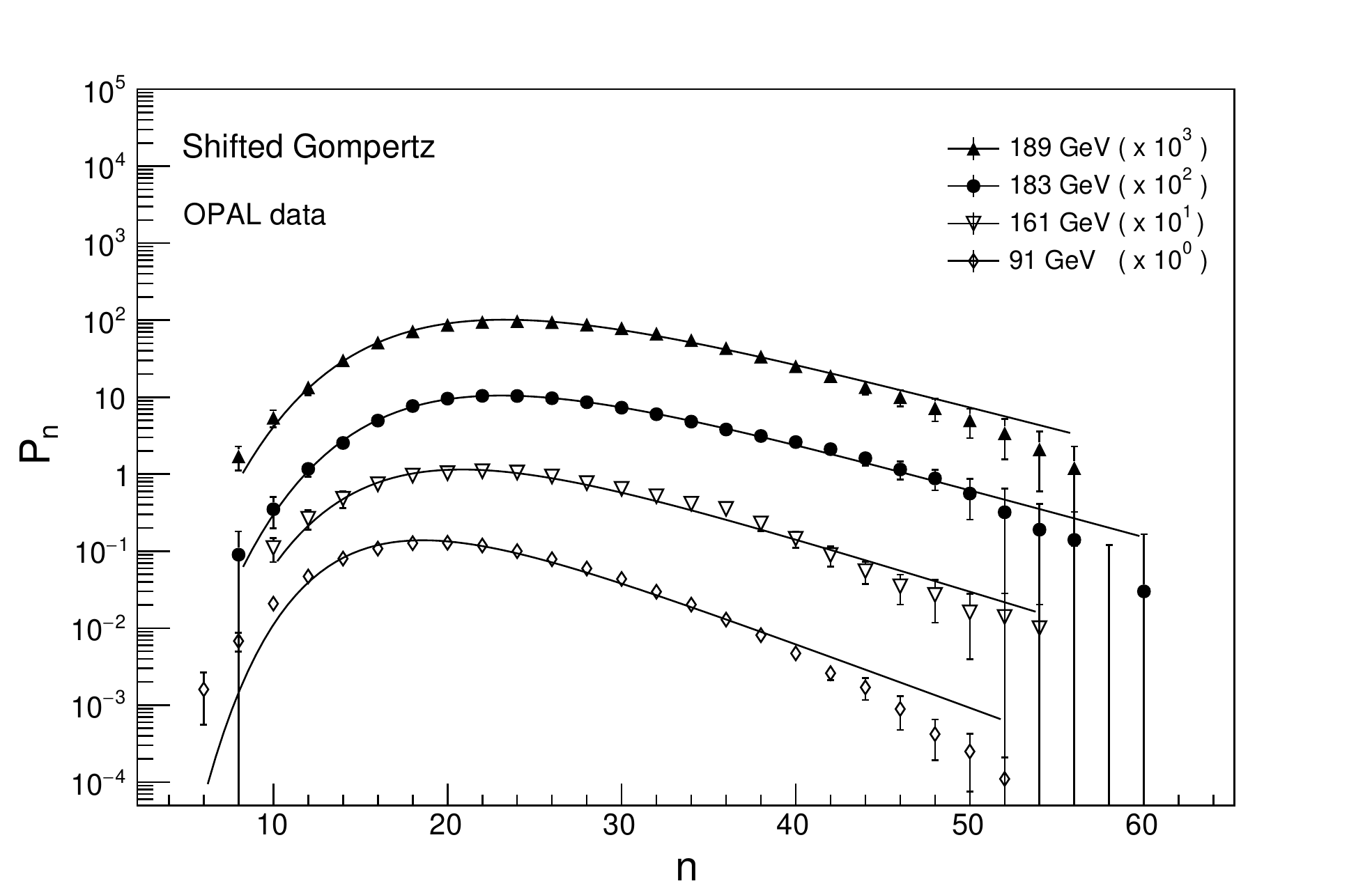}
\includegraphics[width=4.8 in, height =2.8 in]{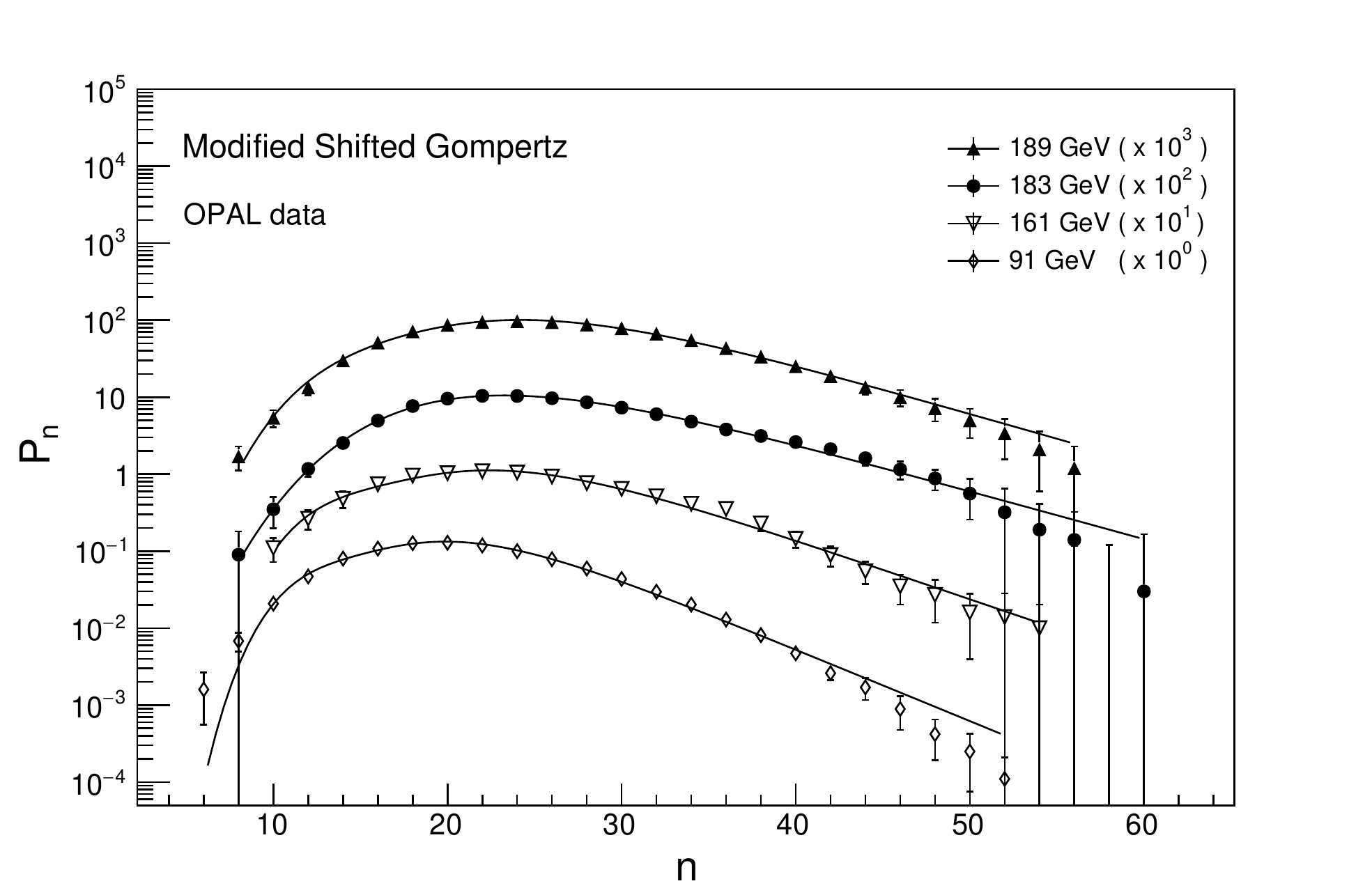}
\caption{Charged multiplicity distribution from OPAL experiment.~Solid lines represent the Gompertz distribution.}
\end{figure}

\begin{figure}[ht]
\includegraphics[width=4.8 in, height =2.8 in]{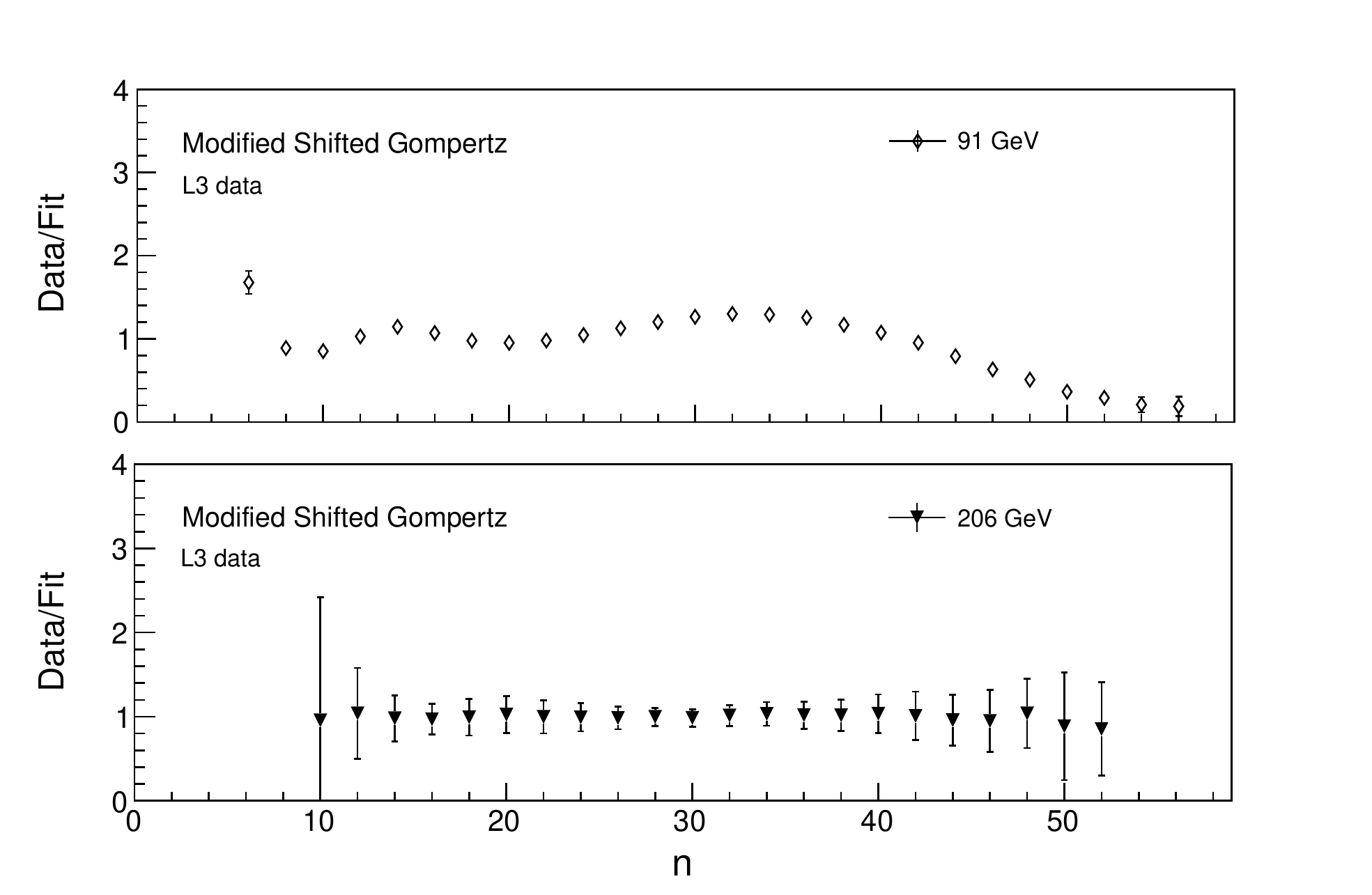}
\includegraphics[width=4.8 in, height =2.8 in]{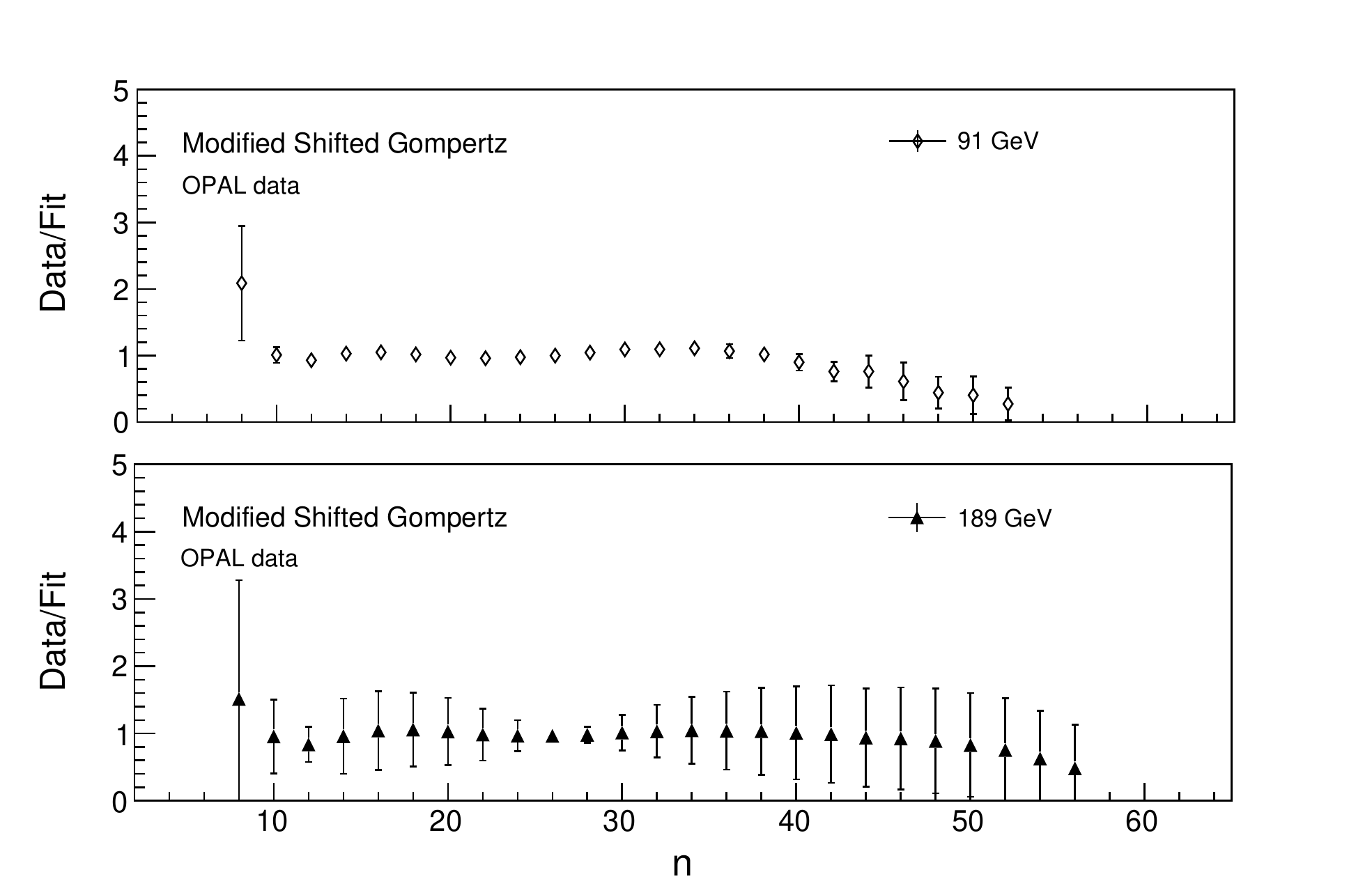}
\caption{Ratio plots of data over modified shifted Gompertz fit values.}
\end{figure}

\begin{table*}[t]
{
\begin{adjustbox}{max width=\textwidth}
\begin{tabular}{|c|c|c|c|c|c|c|c|c|c|c|c|}
\hline
& & & & & & & & & & &\\
& shifted Gompertz & $\rightarrow$ & & & Modified         & $\rightarrow$ & & & & &\\ 
&                  &               & & & shifted Gompertz &               & & & & &\\\hline             
Energy & b & $\eta$ & $\chi^{2}/ndf$ & p value & b$_1$ & $\eta_1$ & b$_2$ & $\eta_2$ & $\alpha$ & $\chi^{2}/ndf$ & p value\\
 (GeV) &   &        &                &         &       &          &       &          &          &                &\\\hline

OPAL & & & & & & & & & & &\\\hline

91 & 0.191 $\pm$ 0.001 & 33.920 $\pm$ 1.133 & 190.90/21 & $<$ 0.0001 & 0.213 $\pm$ 0.003 & 91.100 $\pm$ 7.541 & 0.265 $\pm$ 0.012 & 46.890 $\pm$ 8.189  & 0.657 & 51.88/19 & $<$ 0.0001\\\hline

161 & 0.159 $\pm$ 0.004 & 26.590 $\pm$ 3.120 & 16.72/20 & 0.6711 & 0.178 $\pm$ 0.008 & 67.240 $\pm$ 16.290 & 0.244 $\pm$ 0.040 & 45.310 $\pm$ 24.960 & 0.716 & 7.13/18 & 0.9890\\\hline

183 & 0.142 $\pm$ 0.003 & 25.370 $\pm$ 1.773 & 7.49/24 & 0.9995 & 0.145 $\pm$ 0.010 & 30.810 $\pm$ 6.324  & 0.143 $\pm$ 0.039 & 20.470 $\pm$ 9.522  & 0.675 & 7.06/22 & 0.9989\\\hline

189 & 0.135 $\pm$ 0.002 & 21.930 $\pm$ 1.265 & 22.62/22 & 0.4234 & 0.149 $\pm$ 0.005 & 47.490 $\pm$ 8.789  & 0.177 $\pm$ 0.017 & 25.960 $\pm$ 6.152  & 0.662 & 8.68/20 & 0.9863\\\hline

L3 & & & & & & & & & & &\\\hline

91 & 0.215 $\pm$ 0.001 & 45.170 $\pm$ 0.563 & 3989.00/25 & $<$ 0.0001 & 0.234 $\pm$ 0.001 & 103.100 $\pm$ 1.989 & 0.244 $\pm$ 0.003 & 28.330 $\pm$ 0.870  & 0.651 & 1094.00/23 & $<$ 0.0001\\\hline

131 & 0.173 $\pm$ 0.004 & 31.430 $\pm$ 3.157 & 12.66/15 & 0.6285 & 0.194 $\pm$ 0.009 & 87.480 $\pm$ 27.270 & 0.242 $\pm$ 0.031 & 50.250 $\pm$ 21.510 & 0.654 & 5.32/13 & 0.9675\\\hline

136 & 0.157 $\pm$ 0.004 & 23.140 $\pm$ 2.185 & 27.42/16 & 0.0370 & 0.183 $\pm$ 0.009 & 82.870 $\pm$ 25.990 & 0.258 $\pm$ 0.025 & 56.730 $\pm$ 19.510 & 0.657 & 19.73/14 & 0.1389\\\hline

172 & 0.138 $\pm$ 0.003 & 22.760 $\pm$ 1.427 & 4.82/19 & 0.9996 & 0.146 $\pm$ 0.008 & 36.660 $\pm$ 12.880 & 0.169 $\pm$ 0.055 & 22.680 $\pm$ 16.590 & 0.767 & 2.61/17 & 1.0000\\\hline

182 & 0.138 $\pm$ 0.003 & 23.120 $\pm$ 1.469 & 8.07/19 & 0.9860 & 0.148 $\pm$ 0.008 & 46.180 $\pm$ 12.560 & 0.192 $\pm$ 0.023 & 35.170 $\pm$ 11.900 & 0.668 & 5.25/17 & 0.9970\\\hline

188 & 0.138 $\pm$ 0.002 & 23.860 $\pm$ 1.179 & 26.16/17 & 0.0716 & 0.156 $\pm$ 0.006 & 58.890 $\pm$ 10.230 & 0.199 $\pm$ 0.016 & 38.780 $\pm$ 9.268  & 0.670 & 10.77/15 & 0.7687\\\hline

194 & 0.136 $\pm$ 0.003 & 22.900 $\pm$ 1.479 & 11.83/19 & 0.8928 & 0.149 $\pm$ 0.008 & 25.140 $\pm$ 3.422 & 0.164 $\pm$ 0.016 & 114.500 $\pm$ 67.31 & 0.772 & 6.59/17 & 0.9883\\\hline

200 & 0.132 $\pm$ 0.003 & 22.220 $\pm$ 1.356 & 3.96/19 & 0.9999 & 0.134 $\pm$ 0.009 & 29.340 $\pm$ 12.030 & 0.169 $\pm$ 0.059 & 26.260 $\pm$ 17.710 & 0.779 & 3.73/17 & 0.9997\\\hline

206 & 0.129 $\pm$ 0.003 & 22.320 $\pm$ 1.233 & 1.42/19 & 1.0000 & 0.135 $\pm$ 0.013 & 25.170 $\pm$ 5.682 & 0.114 $\pm$ 0.050 & 14.700 $\pm$ 12.630 & 0.790 & 1.32/17 & 1.0000\\\hline

\end{tabular}
\end{adjustbox}
\caption{Parameters of shifted Gompertz and modified shifted Gompertz functions for $e^+e^-$ collisions.}
}
\end{table*}

\begin{figure}[ht]
\includegraphics[width=4.8 in, height =2.1 in]{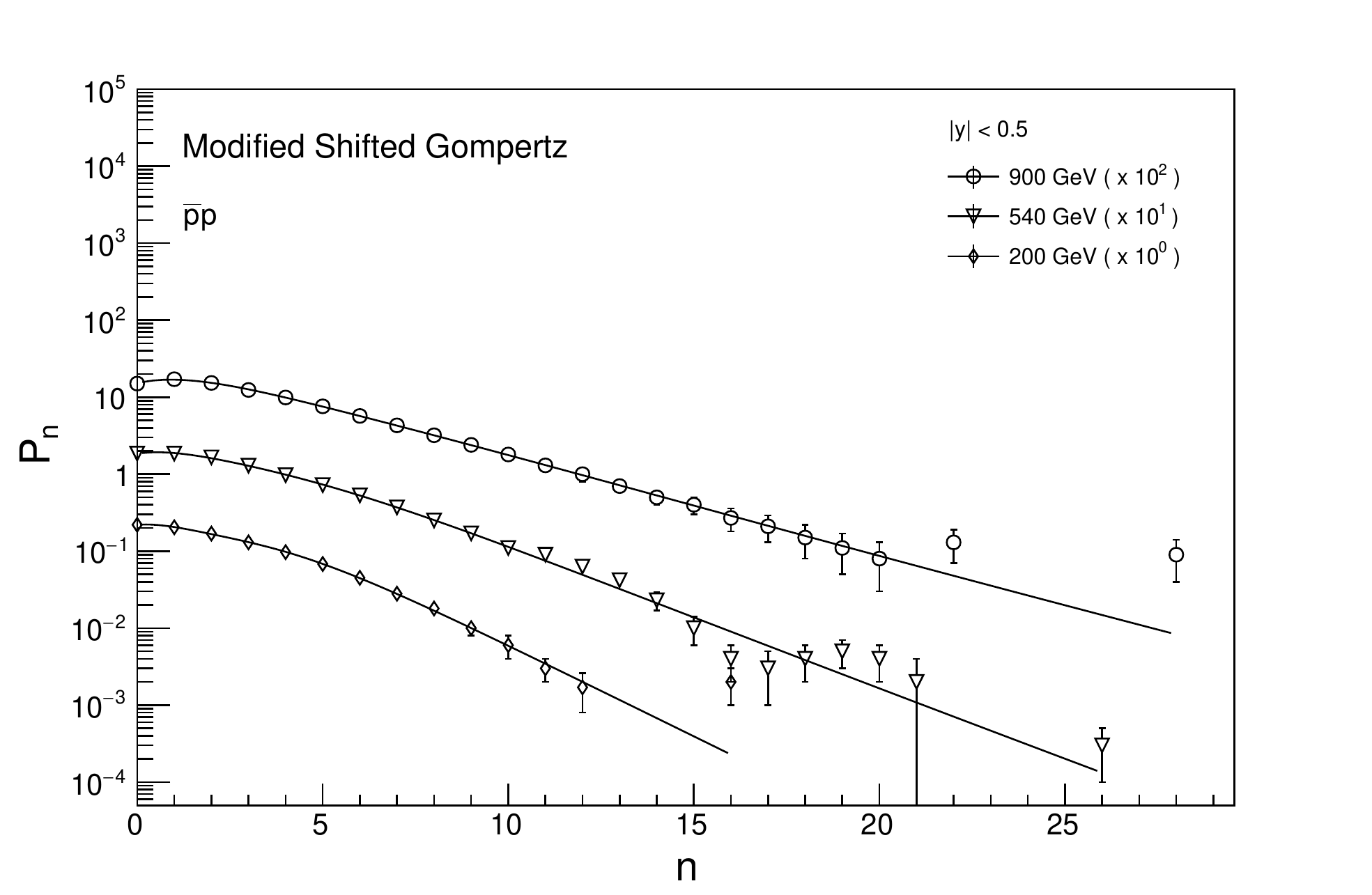}
\includegraphics[width=4.8 in, height =2.1 in]{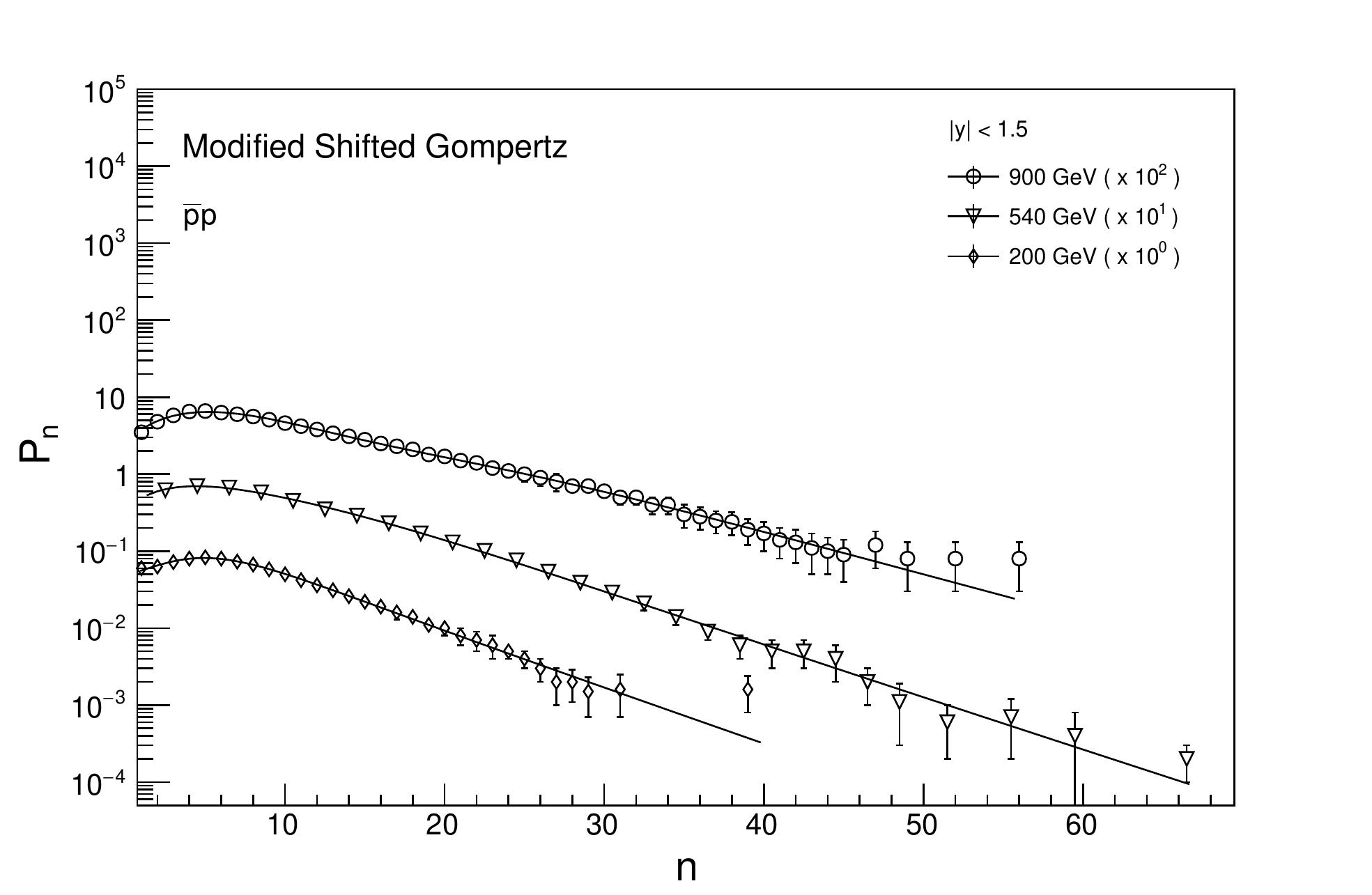}
\includegraphics[width=4.8 in, height =2.1 in]{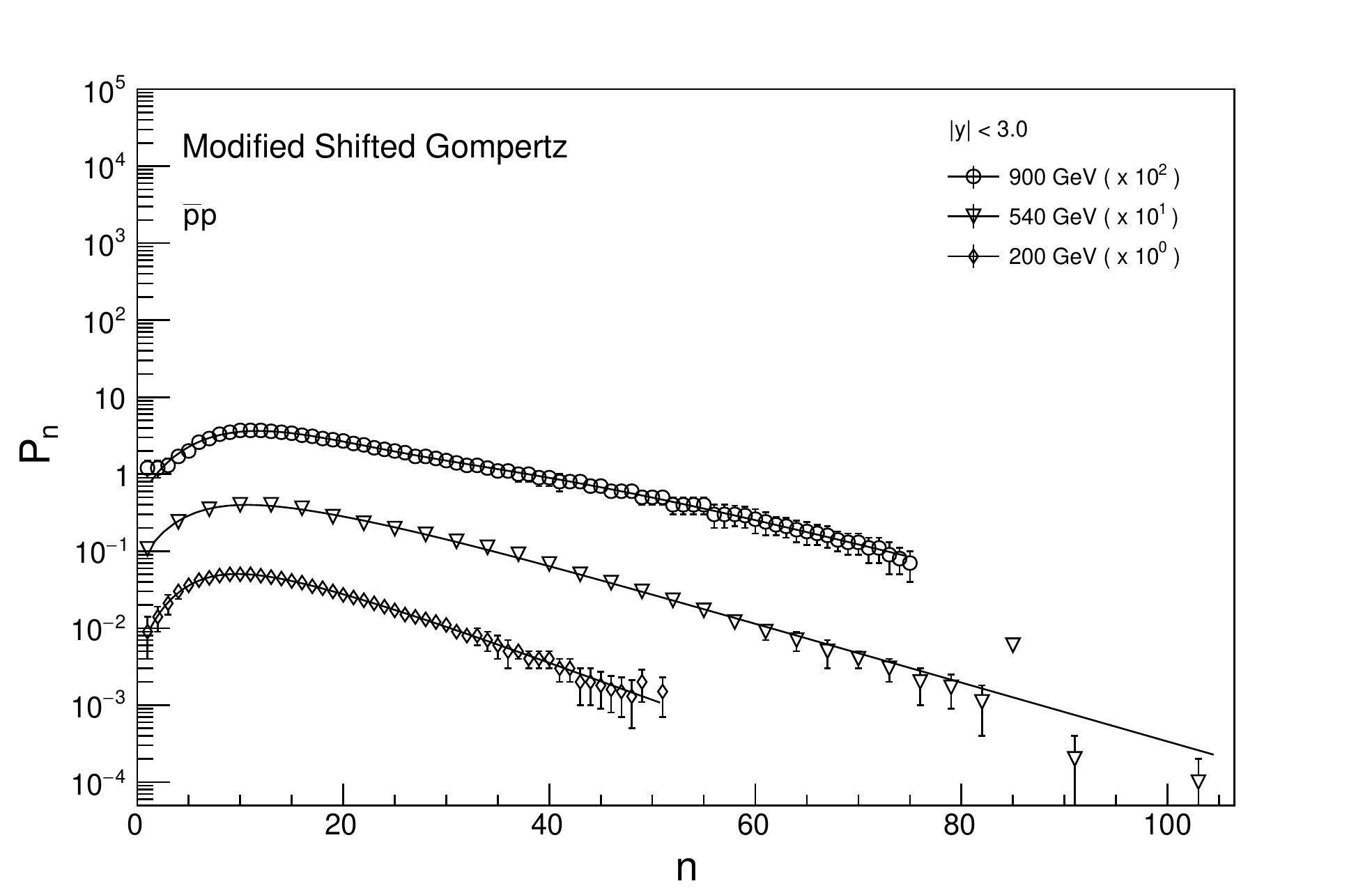}
\includegraphics[width=4.8 in, height =2.1 in]{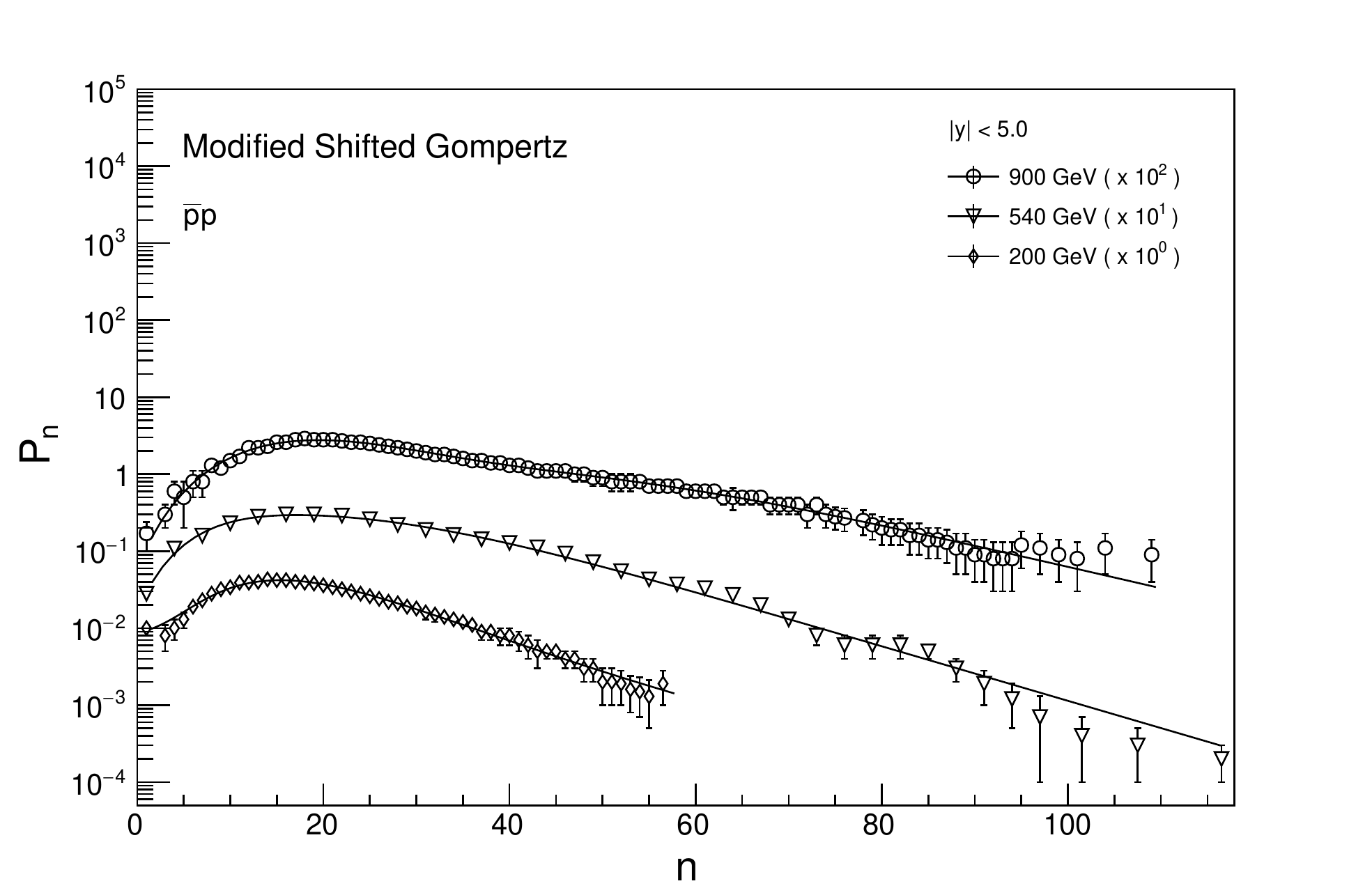}
\caption{Charged multiplicity distributions in $\overline{p}p$ collisions in four rapidity windows. Solid lines represent modified shifted Gompertz function.}
\end{figure}

\begin{figure}[ht]
\includegraphics[width=4.8 in, height =2.8 in]{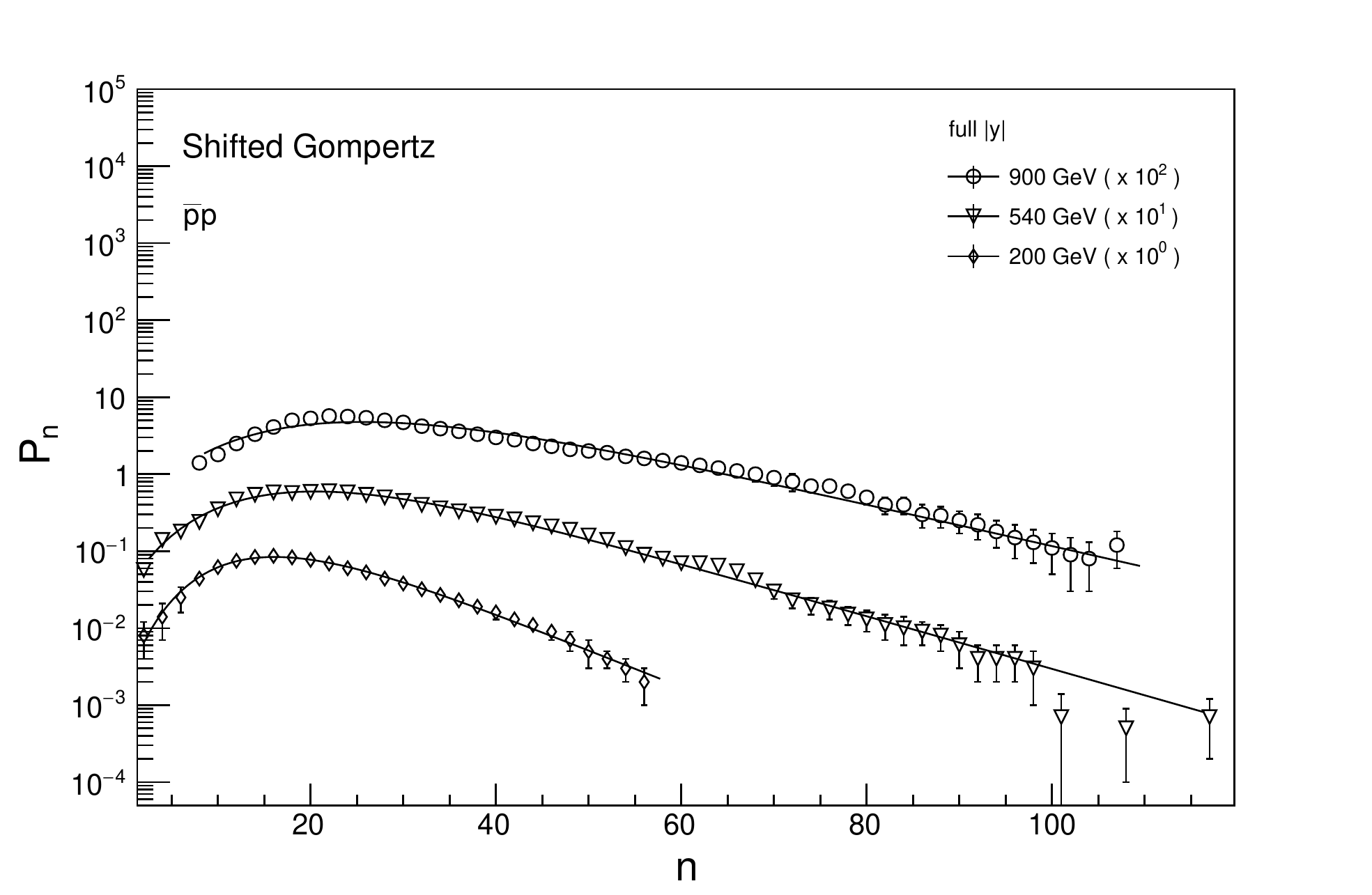}
\includegraphics[width=4.8 in, height =2.8 in]{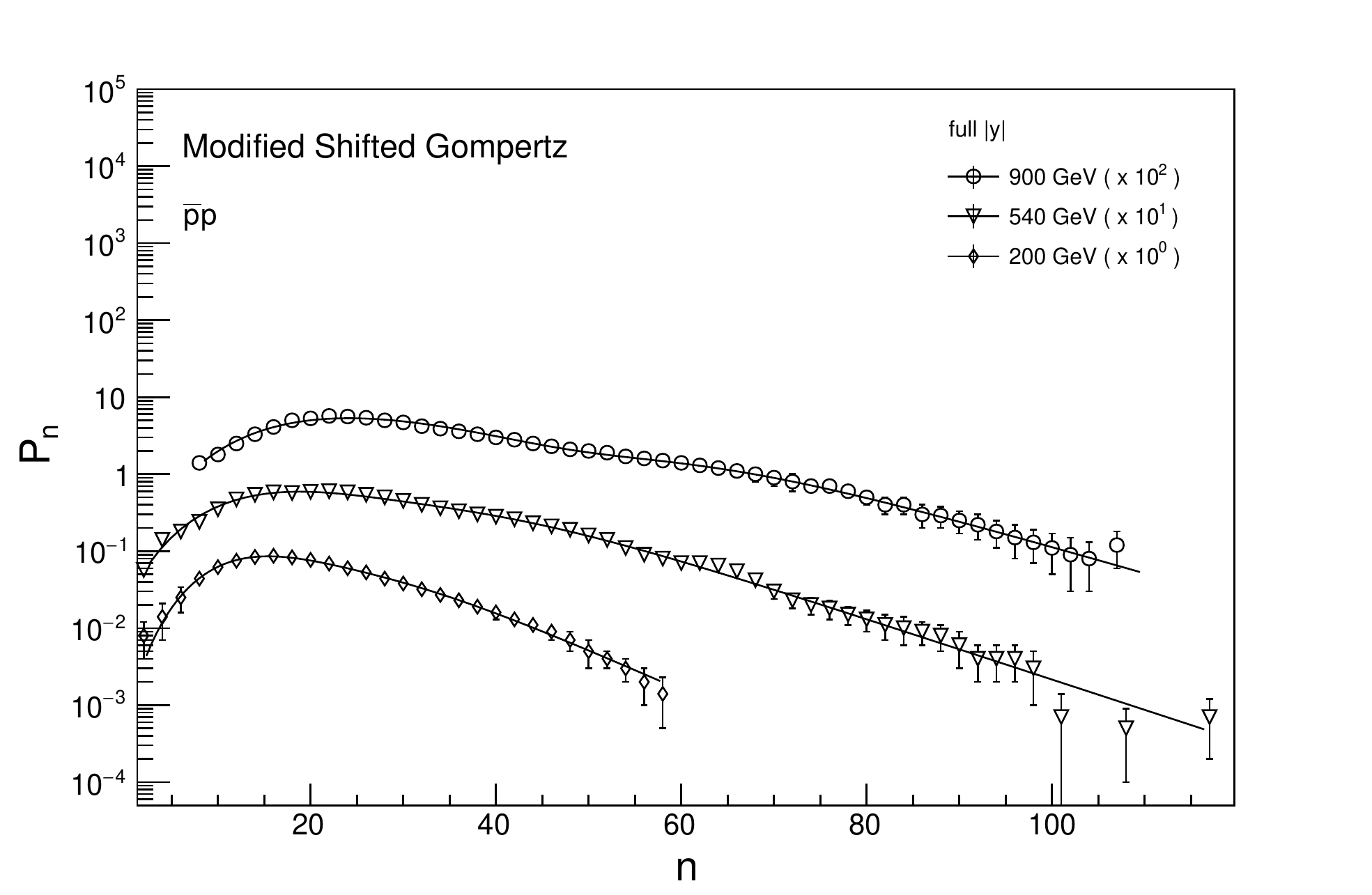}
\caption{Charged multiplicity distributions in $\overline{p}p$ collisions in full phase space.~Solid lines represent shifted Gompertz function (top) and modified shifted Gompertz distributions(bottom).}
\end{figure}

\begin{figure}[ht]
\includegraphics[width=4.8 in, height =2.8 in]{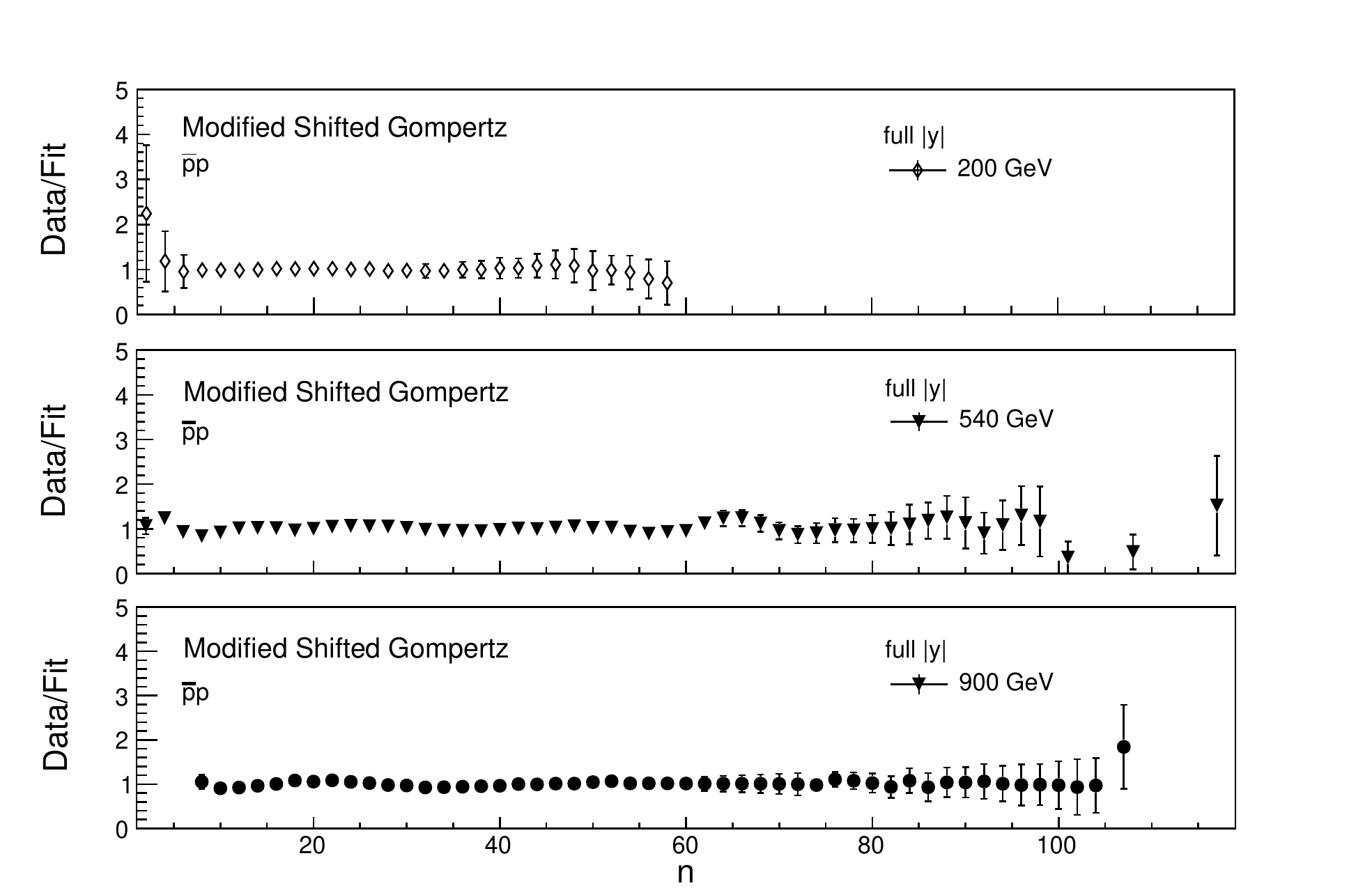}
\caption{Ratio plots of data versus modified shifted Gompertz fit values for $\overline{p}p$ collisions in full phase space.}
\end{figure}

\begin{figure}[ht]
\includegraphics[width=4.8 in, height =2.1 in]{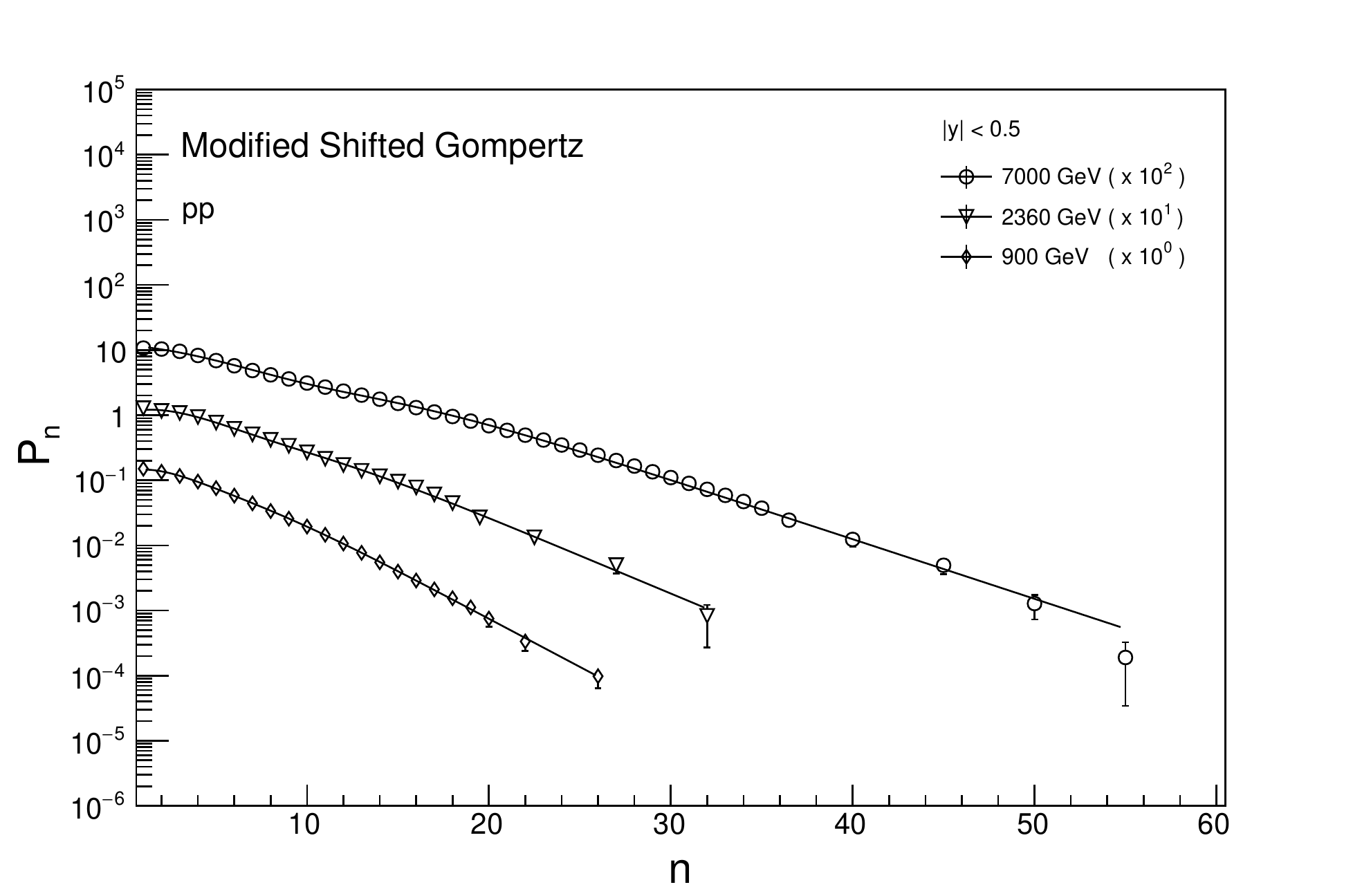}
\includegraphics[width=4.8 in, height =2.1 in]{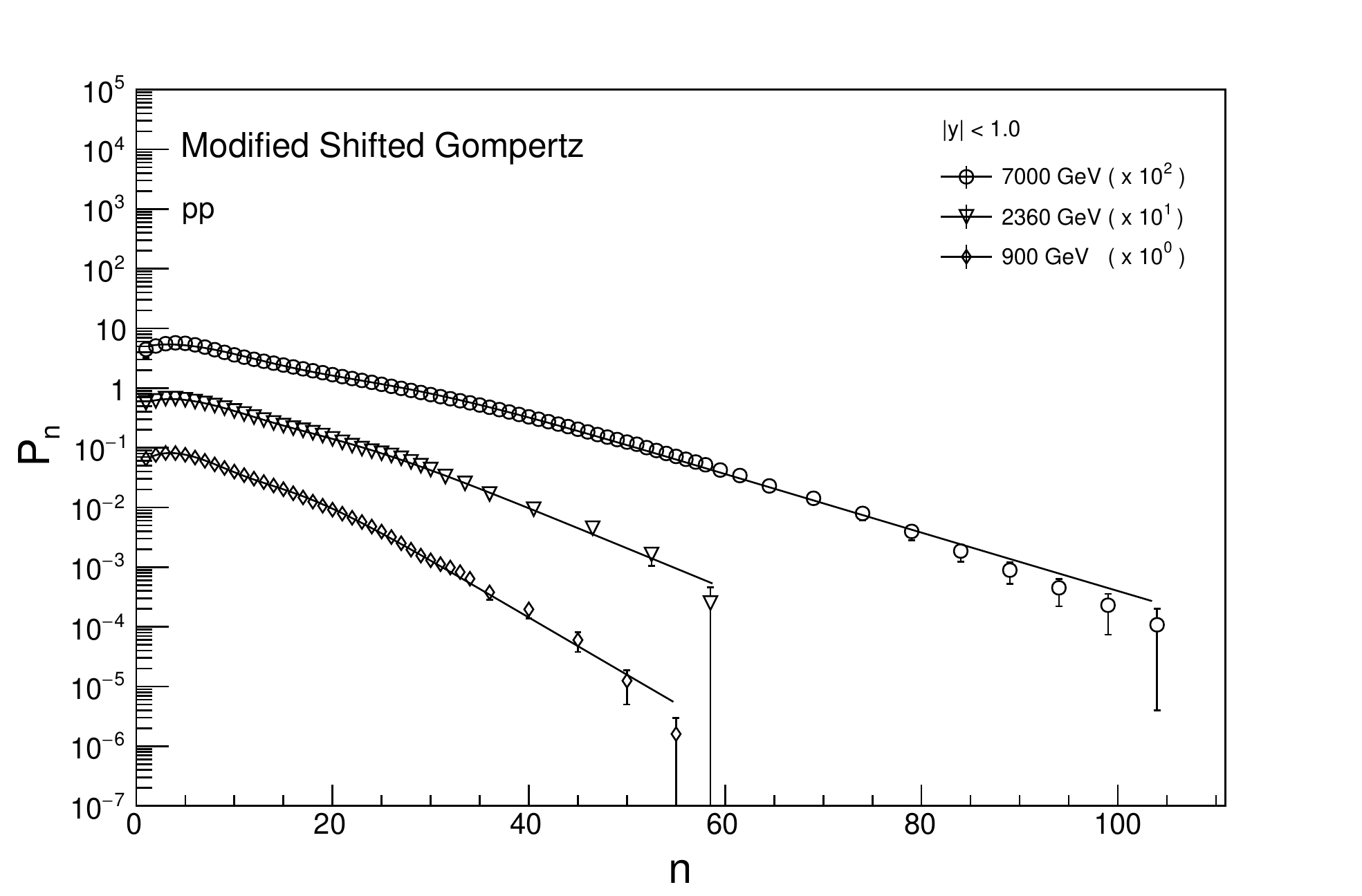}
\includegraphics[width=4.8 in, height =2.1 in]{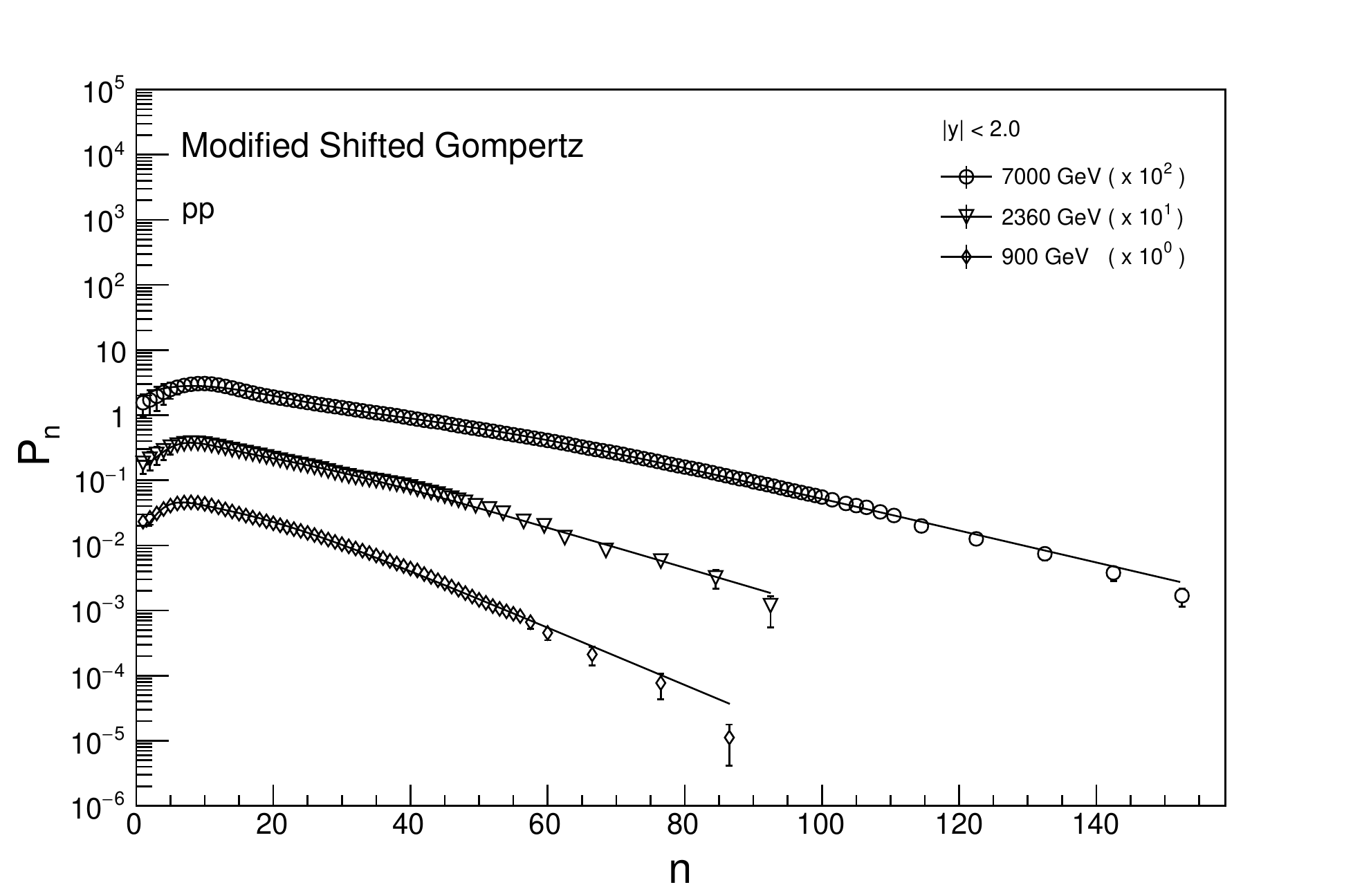}
\includegraphics[width=4.8 in, height =2.1 in]{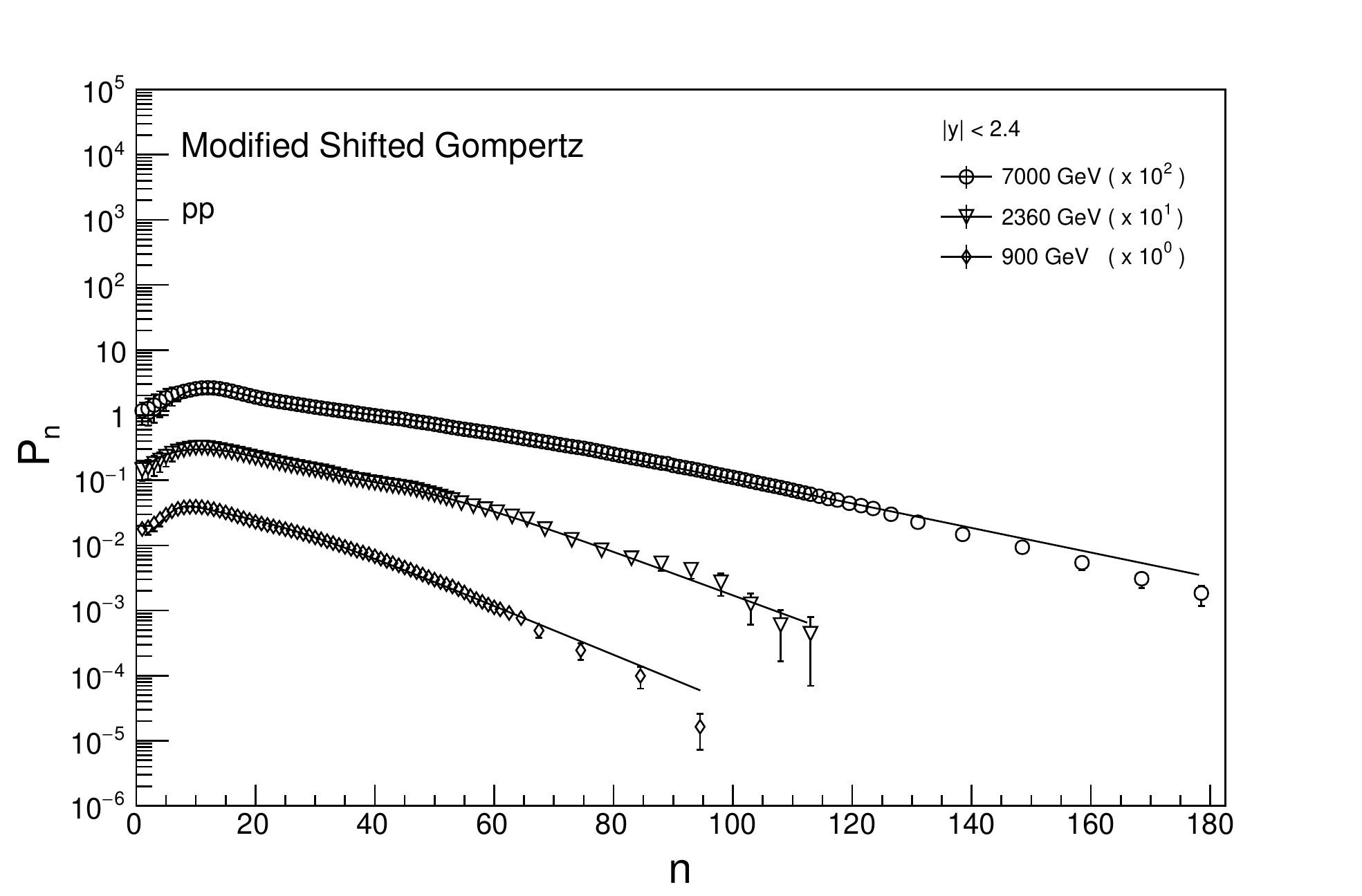}
\caption{Charged multiplicity distributions in $pp$ collisions in four rapidity windows for the modified shifted Gompertz distributions.}
\end{figure}

\begin{figure}[ht]
\includegraphics[width=4.8 in, height=2.8 in]{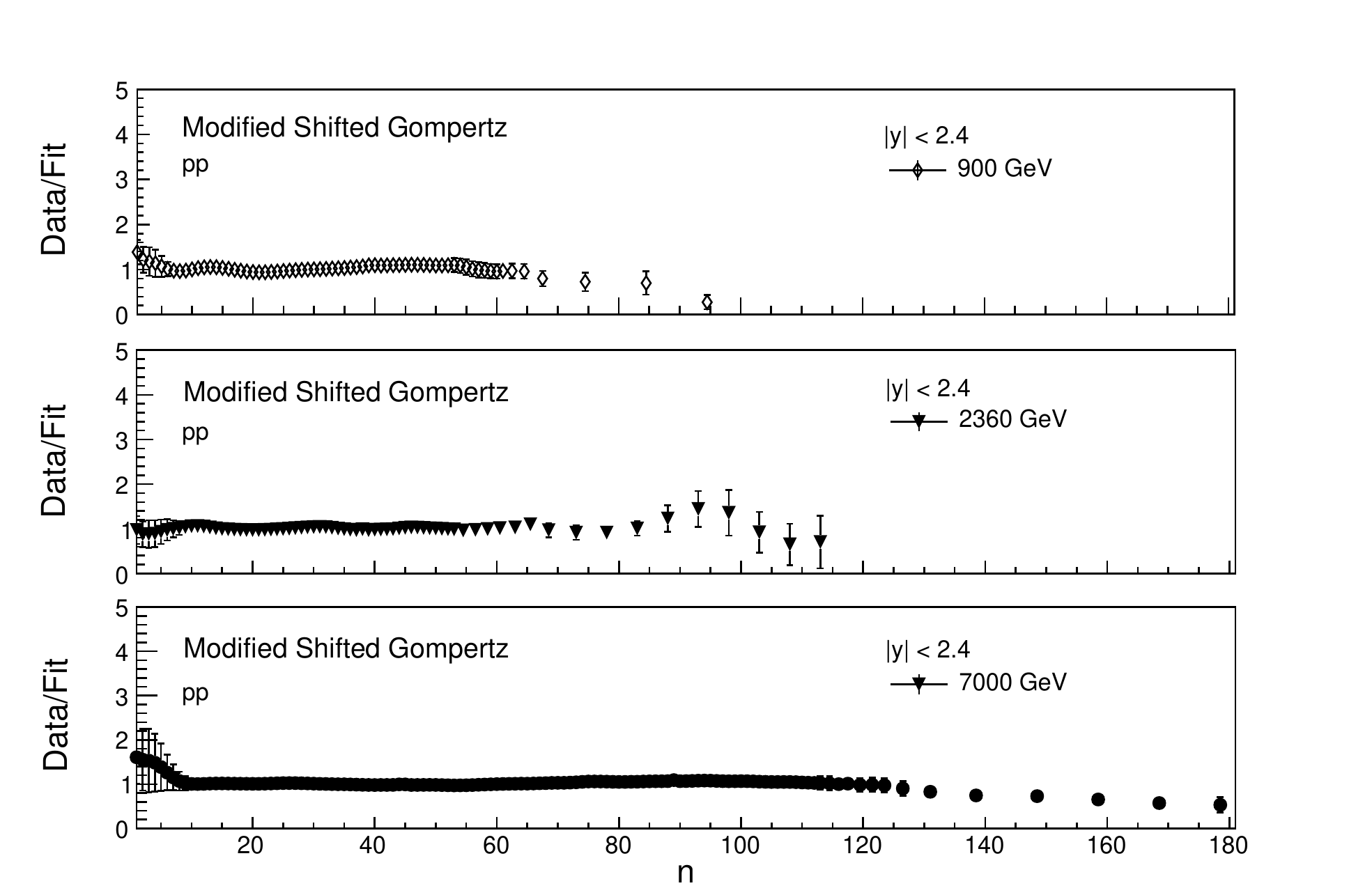}
\caption{ratio plots of data versus modified shifted Gompertz fit values for the $pp$ collisions in full phase space.}
\end{figure}

\begin{table*}[t]
{
\begin{adjustbox}{max width=\textwidth}
\begin{tabular}{|c|c|c|c|c|c|c|c|c|c|c|c|c|}
\hline
& & & & & & & & & & & &\\
& & Shifted Gompertz & $\rightarrow$ & & & Modified & $\rightarrow$ & & & & &\\ 
& &                  &               & & & shifted Gompertz &       & & & & &\\\hline             
Energy & $|y|$ & b & $\eta$ & $\chi^{2}/ndf$ & p value & b$_1$ & $\eta_1$ & b$_2$ & $\eta_2$ & $\alpha$ & $\chi^{2}/ndf$ & p value\\
 (GeV) &       &   &        &                &         &       &          &       &          &          &                &\\\hline

%200 & & & & & & & & & & &\\\hline

200 & 0.5 & 0.455 $\pm$ 0.013 & 0.652 $\pm$ 0.061 & 11.66/11 & 0.3897 & 0.518 $\pm$ 0.041 & 0.584 $\pm$ 0.105 & 0.572 $\pm$ 0.061 & 8.898 $\pm$ 4.092 & 0.850 & 3.79/9 & 0.9247\\\hline

200 & 1.5 & 0.194 $\pm$ 0.005 & 1.570 $\pm$ 0.099 & 9.11/29 & 0.9998 & 0.171 $\pm$ 0.007 & 1.064 $\pm$ 0.127 & 0.338 $\pm$ 0.052 & 6.284 $\pm$ 2.489 & 0.812 & 5.39/27 & 1.0000\\\hline

200 & 3.0 & 0.123 $\pm$ 0.003 & 2.606 $\pm$ 0.179 & 12.60/47 & 1.0000 & 0.159 $\pm$ 0.010 & 3.455 $\pm$ 0.332 & 0.108 $\pm$ 0.015 & 9.180 $\pm$ 5.148 & 0.800 & 2.98/45 & 1.0000\\\hline

200 & 5.0 & 0.101 $\pm$ 0.002 & 3.381 $\pm$ 0.175 & 35.33/52 & 0.9628 & 0.133 $\pm$ 0.013 & 4.921 $\pm$ 0.774 & 0.097 $\pm$ 0.016 & 10.430 $\pm$ 7.851 & 0.760 & 3.48/46 & 1.0000\\\hline

200 & full & 0.111 $\pm$ 0.003 & 5.033 $\pm$ 0.353 & 3.96/25 & 1.0000 & 0.107 $\pm$ 0.004 & 4.830 $\pm$ 0.641 & 0.197 $\pm$ 0.054 & 15.400 $\pm$ 12.430 & 0.900 & 4.59/26 & 1.0000\\\hline

540 & 0.5 & 0.397 $\pm$ 0.005 & 0.682 $\pm$ 0.033 & 26.90/20 & 0.1381 & 0.490 $\pm$ 0.029 & 0.674 $\pm$ 0.074 & 0.407 $\pm$ 0.019 & 5.363 $\pm$ 1.598 & 0.810 & 21.29/18 & 0.2650\\\hline

540 & 1.5 & 0.162 $\pm$ 0.002 & 1.387 $\pm$ 0.049 & 17.22/26 & 0.0370 & 0.224 $\pm$ 0.012 & 1.504 $\pm$ 0.109 & 0.155 $\pm$ 0.006 & 5.071 $\pm$ 1.088 & 0.700 & 11.04/24 & 0.9887\\\hline

540 & 3.0 & 0.097 $\pm$ 0.001 & 2.308 $\pm$ 0.055 & 176.40/28 & $<$ 0.0001 & 0.089 $\pm$ 0.003 & 3.070 $\pm$ 0.389 & 0.158 $\pm$ 0.008 & 2.794 $\pm$ 0.205 & 0.640 & 21.29/23 & 0.5634\\\hline

540 & 5.0 & 0.080 $\pm$ 0.001 & 3.489 $\pm$ 0.069 & 69.33/33 & 0.0002 & 0.081 $\pm$ 0.001 & 7.072 $\pm$ 0.566 & 0.126 $\pm$ 0.005 & 3.579 $\pm$ 0.165 & 0.580 & 48.50/31 & 0.0236\\\hline

540 & full & 0.079 $\pm$ 0.001 & 4.088 $\pm$ 0.094 & 59.83/49 & 0.0226 & 0.079 $\pm$ 0.002 & 5.932 $\pm$ 0.649 & 0.116 $\pm$ 0.012 & 3.633 $\pm$ 0.345 & 0.730 & 58.97/47 & 0.1130\\\hline

900 & 0.5 & 0.327 $\pm$ 0.008 & 0.616 $\pm$ 0.063 & 10.16/20 & 0.9652 & 0.431 $\pm$ 0.033 & 0.798 $\pm$ 0.116 & 0.284 $\pm$ 0.041 & 4.561 $\pm$ 2.874 & 0.845 & 5.13/18 & 0.9986\\\hline

900 & 1.5 & 0.129 $\pm$ 0.003 & 1.083 $\pm$ 0.087 & 35.85/46 & 0.8593 & 0.191 $\pm$ 0.007 & 1.834 $\pm$ 0.155 & 0.127 $\pm$ 0.011 & 12.530 $\pm$ 3.698 & 0.841 & 5.77/44 & 1.0000\\\hline

900 & 3.0 & 0.076 $\pm$ 0.002 & 1.834 $\pm$ 0.103 & 59.90/71 & 0.8234 & 0.129 $\pm$ 0.005 & 3.256 $\pm$ 0.274 & 0.075 $\pm$ 0.004 & 10.320 $\pm$ 2.117 & 0.725 & 8.01/69 & 1.0000\\\hline

900 & 5.0 & 0.063 $\pm$ 0.001 & 3.226 $\pm$ 0.125 & 89.95/95 & 0.6272 & 0.104 $\pm$ 0.004 & 5.355 $\pm$ 0.492 & 0.057 $\pm$ 0.003 & 10.370 $\pm$ 2.132 & 0.660 & 17.98/93 & 1.0000\\\hline

900 & full & 0.063 $\pm$ 0.001 & 4.043 $\pm$ 0.195 & 67.16/47 & 0.0283 & 0.101 $\pm$ 0.003 & 7.852 $\pm$ 0.618 & 0.061 $\pm$ 0.003 & 18.170 $\pm$ 3.879 & 0.710 & 13.46/45 & 1.0000\\\hline

\end{tabular}
\end{adjustbox}
\caption{Parameters of shifted Gompertz and modified shifted Gompertz functions for $\overline{p}p$ collisions}
}
\end{table*}

Figure~4 shows the modified shifted Gompertz distribution, equation~(6) fitted to the $\overline{p}p$ data at energies from 200 GeV to 900 GeV in four rapidity windows.~To avoid cluttering of figures, the plots for shifted Gompertz are not shown.~Figure~5 shows the shifted Gompertz and modified shifted Gompertz distributions, fitted to the $\overline{p}p$ collisions in full phase space for the same energies.~The comparison can be seen from the parameters of the fits, $\chi^{2}/ndf$ and the p-values  documented in Table~II.~Figure~6 shows the ratio plots of the data over modified shifted Gompertz fit for  $\overline{p}p$ collisions at different energies in full phase space.~The plots show acceptable fluctuations with the ratio values around unity.

\begin{table*}[t]
{\
\begin{adjustbox}{max width=\textwidth}
\begin{tabular}{|c|c|c|c|c|c|c|c|c|c|c|c|c|}
\hline
& & & & & & & & & & & &\\
& & Shifted Gompertz & $\rightarrow$ & & & Modified & $\rightarrow$ & & & & &\\ 
& &                  &               & & & Shifted Gompertz &       & & & & &\\\hline             
Energy & $|y|$ & b & $\eta$ & $\chi^{2}/ndf$ & p value & b$_1$ & $\eta_1$ & b$_2$ & $\eta_2$ & $\alpha$ & $\chi^{2}/ndf$ & p value\\
 (GeV) &       &   &        &                &         &       &          &       &          &          &                &\\\hline

%200 & & & & & & & & & & &\\\hline

900 & 0.5 & 0.320 $\pm$ 0.005 & 0.719 $\pm$ 0.113 & 3.57/19 & 1.0000 & 0.327 $\pm$ 0.007 & 1.250 $\pm$ 0.243 & 0.949 $\pm$ 0.370 & 1.259 $\pm$ 2.089 & 0.840 & 2.24/17 & 1.0000\\\hline

900 & 1.0 & 0.193 $\pm$ 0.002 & 1.307 $\pm$ 0.109 & 84.97/36 & $<$ 0.0001 & 0.198 $\pm$ 0.002 & 2.260 $\pm$ 0.159 & 0.667 $\pm$ 0.079 & 3.366 $\pm$ 1.286 & 0.810 & 63.32/34 & 0.0017\\\hline

900 & 1.5 & 0.131 $\pm$ 0.001 & 1.320 $\pm$ 0.092 & 66.98/48 & 0.0364 & 0.136 $\pm$ 0.002 & 2.553 $\pm$ 0.169 & 0.397 $\pm$ 0.036 & 3.114 $\pm$ 0.866 & 0.760 & 44.22/46 & 0.5471\\\hline

900 & 2.0 & 0.101 $\pm$ 0.001 & 1.431 $\pm$ 0.084 & 55.41/58 & 0.5722 & 0.104 $\pm$ 0.001 & 2.660 $\pm$ 0.169 & 0.283 $\pm$ 0.024 & 3.123 $\pm$ 0.756 & 0.750 & 33.43/56 & 0.9928\\\hline

900 & 2.4 & 0.087 $\pm$ 0.001 & 1.585 $\pm$ 0.081 & 72.26/64 & 0.2239 & 0.088 $\pm$ 0.001 & 2.662 $\pm$ 0.162 & 0.250 $\pm$ 0.021 & 3.816 $\pm$ 1.020 & 0.780 & 48.03/62 & 0.9036\\\hline

2360 & 0.5 & 0.242 $\pm$ 0.005 & 0.516 $\pm$ 0.099 & 8.13/19 & 0.9853 & 0.424 $\pm$ 0.042 & 1.148 $\pm$ 0.433 & 0.253 $\pm$ 0.011 & 5.447 $\pm$ 1.320 & 0.720 & 5.70/17 & 0.9950\\\hline

2360 & 1.0 & 0.133 $\pm$ 0.002 & 0.700 $\pm$ 0.091 & 24.30/34 & 0.8904 & 0.139 $\pm$ 0.003 & 2.472 $\pm$ 0.270 & 0.352 $\pm$ 0.033 & 1.823 $\pm$ 0.550 & 0.620 & 14.03/32 & 0.9975\\\hline

2360 & 1.5 & 0.092 $\pm$ 0.002 & 0.819 $\pm$ 0.090 & 28.08/45 & 0.9773 & 0.099 $\pm$ 0.002 & 3.089 $\pm$ 0.303 & 0.249 $\pm$ 0.019 & 2.195 $\pm$ 0.499 & 0.590 & 10.03/43 & 1.0000\\\hline

2360 & 2.0 & 0.071 $\pm$ 0.001 & 0.917 $\pm$ 0.088 & 39.83/55 & 0.9383 & 0.076 $\pm$ 0.002 & 3.342 $\pm$ 0.321 & 0.190 $\pm$ 0.013 & 2.526 $\pm$ 0.487 & 0.580 & 12.70/53 & 1.0000\\\hline

2360 & 2.4 & 0.062 $\pm$ 0.001 & 1.122 $\pm$ 0.089 & 59.55/66 & 0.6993 & 0.137 $\pm$ 0.006 & 2.699 $\pm$ 0.379 & 0.070 $\pm$ 0.002 & 8.042 $\pm$ 0.832 & 0.610 & 14.49/64 & 1.0000\\\hline

7000 & 0.5 & 0.184 $\pm$ 0.002 & 0.580 $\pm$ 0.073 & 117.50/37 & $<$ 0.0001 & 0.387 $\pm$ 0.013 & 1.376 $\pm$ 0.248 & 0.202 $\pm$ 0.003 & 7.402 $\pm$ 0.599 & 0.710 & 24.08/35 & 0.9178\\\hline

7000 & 1.0 & 0.101 $\pm$ 0.001 & 0.846 $\pm$ 0.068 & 223.70/66 & $<$ 0.0001 & 0.229 $\pm$ 0.007 & 1.778 $\pm$ 0.250 & 0.110 $\pm$ 0.001 & 6.826 $\pm$ 0.409 & 0.650 & 48.85/64 & 0.9195\\\hline

7000 & 1.5 & 0.068 $\pm$ 0.001 & 0.854 $\pm$ 0.062 & 247.90/88 & $<$ 0.0001 & 0.192 $\pm$ 0.006 & 2.419 $\pm$ 0.336 & 0.074 $\pm$ 0.001 & 4.459 $\pm$ 0.241 & 0.520 & 60.87/86 & 0.9817\\\hline

7000 & 2.0 & 0.049 $\pm$ 0.001 & 0.667 $\pm$ 0.053 & 164.60/108 & 0.0004 & 0.136 $\pm$ 0.005 & 2.111 $\pm$ 0.276 & 0.055 $\pm$ 0.001 & 4.306 $\pm$ 0.229 & 0.530 & 27.79/106 & 1.0000\\\hline

7000 & 2.4 & 0.042 $\pm$ 0.001 & 0.693 $\pm$ 0.051 & 179.70/123 & 0.0007 & 0.046 $\pm$ 0.001 & 3.871 $\pm$ 0.195 & 0.122 $\pm$ 0.004 & 2.396 $\pm$ 0.304 & 0.510 & 30.92/121 & 1.0000\\\hline

\end{tabular}
\end{adjustbox}
\caption{Parameters of shifted Gompertz and modified shifted Gompertz functions for $pp$ collisions.}
}
\end{table*} 

Figure~7 shows the modified shifted Gompertz distribution, equation~(6) fitted to the $pp$ collisions at LHC energies from 900 GeV to 7000 GeV in four rapidity windows.~Again for restricting the number of figures, only the modified distributions are shown.~Comparison between the two types of distributions can be seen from the parameters of the fits, $\chi^{2}/ndf$ and the p-values documented in Table~III.~Figure~8 shows the ratio plots of data over modified shifted Gompertz fit for the collisions  at different energies in full phase space.~The plots show acceptable fluctuations with the ratio values around unity.

Comparison of the fits and the parameters shows that overall shifted Gompertz distribution is able to reproduce the data at most of the collision energies in full phase space as well as in the restricted rapidity windows for $e^{+}e^{-}$, $\overline{p}p$ and $pp$ collisions.~It does fail and is excluded statistically for some energies where p-value is $< 0.1\%$, in particular for 540 GeV $\overline{p}p$ data for some rapidity intervals and for LHC data at the highest energy of 7000 GeV.~However, comparison of the fits and the parameters from the (two-component) modified shifted Gompertz distribution shows that though the data are very well reproduced in full phase space as well as in all rapidity intervals for all collision energies in $e^{+}e^{-}$, $\overline{p}p$ and $pp$ collisions, the distribution does fail for the $e^{+}e^{-}$ collisions at 91 GeV.~The $\chi^{2}/ndf$ value in each case reduces enormously, when modified shifted Gompertz fit is used.~In each case the fit is accepted with p-value $ >0.1\%$.

For shifted Gompertz distribution, the scale parameter $b$ and the shape parameter $\eta$ values are plotted in figure~9 for $e^{+}e^{-}$ interactions for LEP data from L3 and OPAL experiments.~A power law is fitted to the data.~It is observed that both $b$ and $\eta$ values decrease with increase in collision energy and are parametrised as;
\begin{equation}
 b = (1.514 \pm 0.084) \sqrt{s}^{(-0.459 \pm 0.012)}
\end{equation}
\begin{equation} 
\eta = (357.693 \pm 96.837) \sqrt{s}^{(-0.524 \pm 0.053)}
\end{equation}

For minimisation of $\chi^{2}$ for the fits, CERN library MINUIT2 has been used.~In case of modified shifted Gompertz, the fit parameters are doubled while introducing the modification.~This causes large error limits on the parameters resulting in the very large $p$ values, particularly close to 1.~In addition, the LEP data for $e^{+}e^{-}$ collisions suffer from very small sample size at some energies, thereby adding to the errors on the fit parameters. 

Using shifted Gompertz distribution, the multiplicity distribution for 500 GeV $e^{+}e^{-}$ collisions at a future Collider is predicted, as shown in figure~10.~The value of mean multiplicity $<n>$ is predicted to be the 37.14 $\pm$ 1.12.~Figure~11 shows the dependence of mean multiplicity from experimental data on energy $\sqrt{s}$.~The fitted curve in equation (10) represents Fermi-Landau model \cite{Fermi1, Fermi2} and fits the data reasonably well with $a_{1}$ = -10.609 $\pm$ 2.003 and $b_{1}$ = 10.156 $\pm$ 0.561 
\begin{equation}
 <n> = a_{1} + b_{1}\sqrt{s}^{1/4}
\end{equation}
~It may be observed that the value of $<n>$ predicted from shifted Gompertz distribution at 500 GeV fits well on the curve, as shown in the figure~11.~A parameterization of the multiplicity data in $e^+e^-$ collisions at the next-to-leading-order QCD was done by D.E. Groom et al \cite{Groom, Diso} and is given in equation (9) of the reference:

\begin{equation}
\small
<n(s)> = a.exp\left[\frac{4}{\beta_{0}}\sqrt{\frac{6\pi}{\alpha_{s}(s)}} + \left(\frac{1}{4}+\frac{10n_{f}}{27\beta_{0}}\right)ln\alpha_{s}(s)\right] + c
\end{equation}
\normalsize
where $a$ and $c$ are constants and $\beta_{0}$ is defined in equation (9.4b). ~The $<n>$ versus $\sqrt{s}$ dependence was shown in reference \cite{Nason}.~Parameters $a$ and $c$ were fitted to the experimental data and a very good agreement was shown.~It is observed that both formulae, equations (10, 11), provide excellent extrapolations for $\sqrt{s} > 206$.~The mean multiplicity $<n>$ at 500 GeV is predicted to be 39.18 by NLO QCD equation.~In the present work, the mean value predicted by the shifted Gompertz distribution as 37.14 $\pm$ 1.23 agrees very closely with the value derived from NLO QCD.~This is good test of the validity of the proposed distribution. 

\begin{figure}[ht]
\includegraphics[width=4.8 in, height =2.8 in]{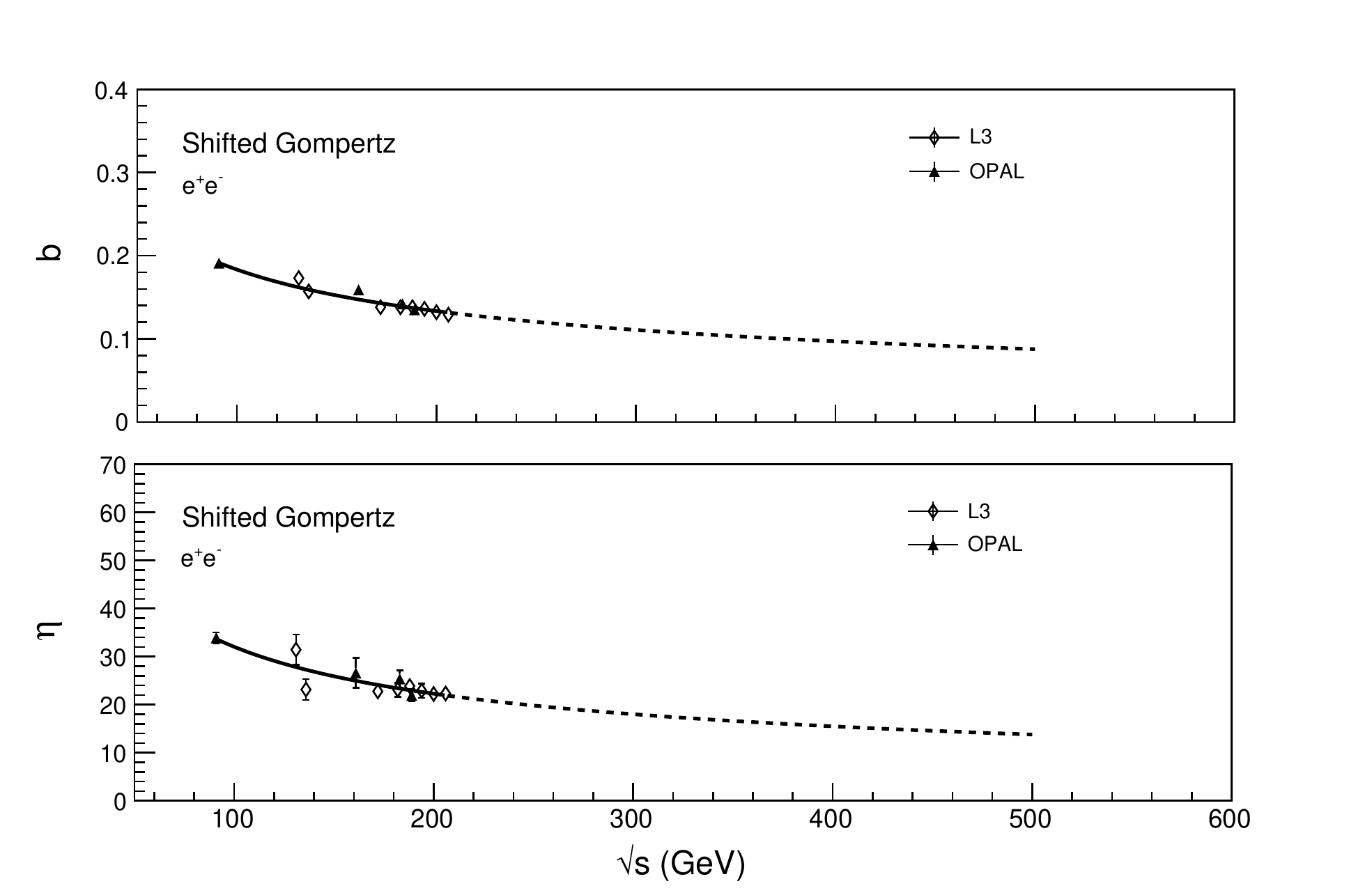}
\caption{Scale parameter $b$ and shape parameter $\eta$ for $e^+e^-$ data.~Points represent the data, solid line is the power law fit and the dotted line the extension of power law fit.}
\end{figure}
\begin{figure}[ht]
\includegraphics[width=4.8 in, height =2.8 in]{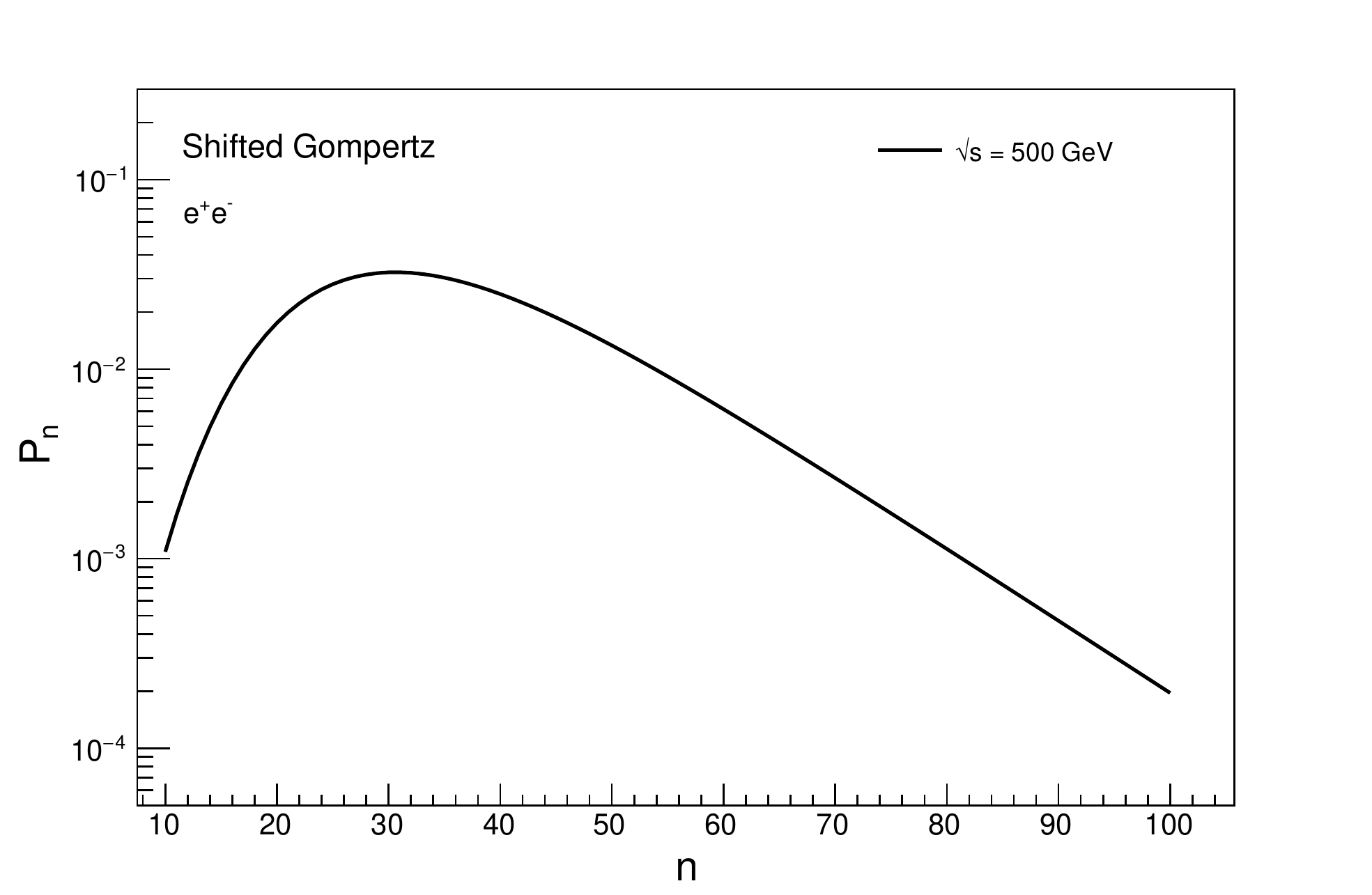}
\caption{Probability distribution for $e^+e^-$ collisions predicted from shifted Gompertz function.}
\end{figure}
\begin{figure}[ht]
\includegraphics[width=4.8 in, height =2.8 in]{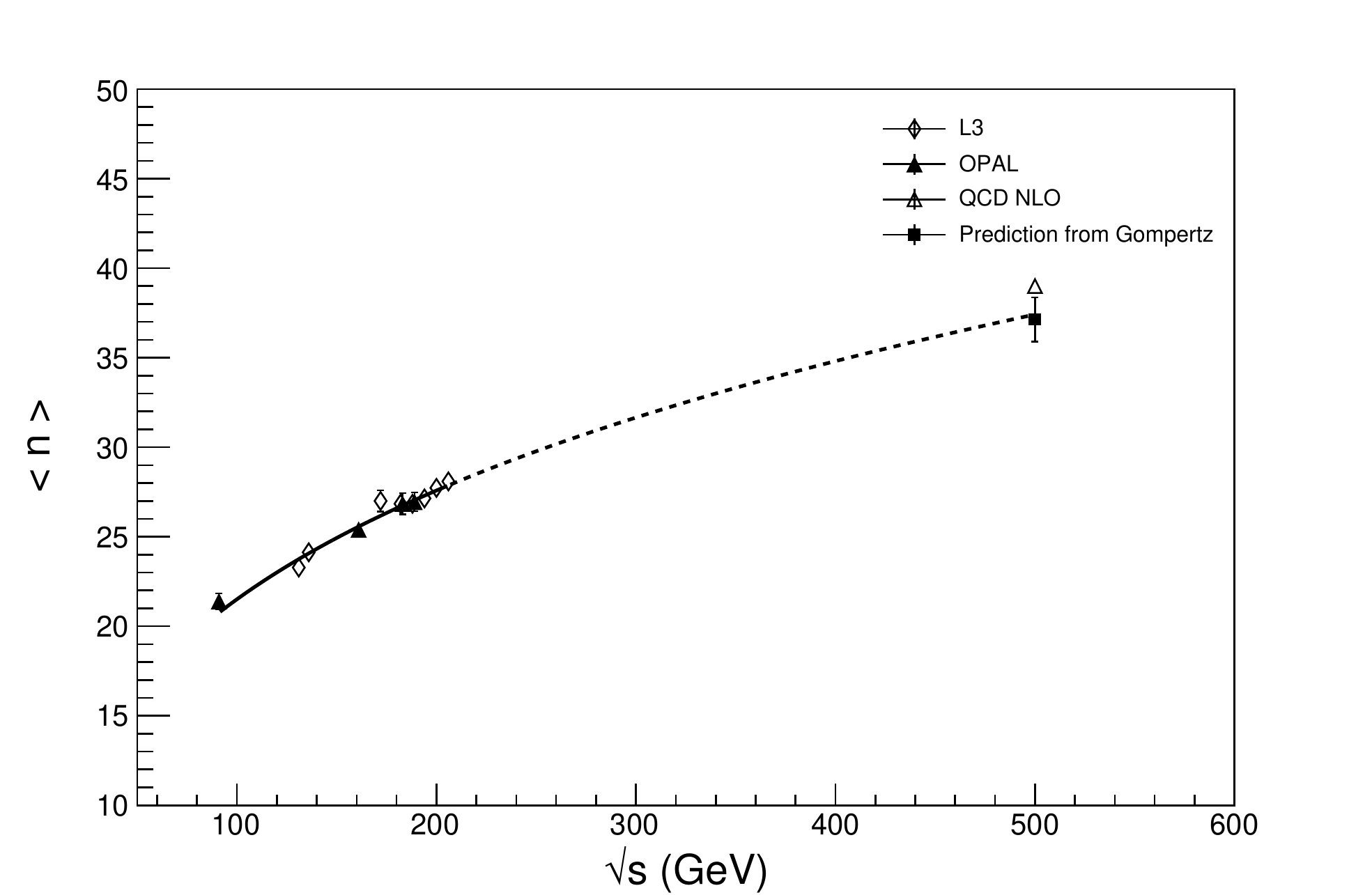}
\caption{Average multiplicity in $e^+e^-$ collisions as a function of $\sqrt{s}$.~The fitted curve represents Fermi-Landau model.}
\end{figure}

\begin{figure}[ht]
\includegraphics[width=4.8 in, height =2.8 in]{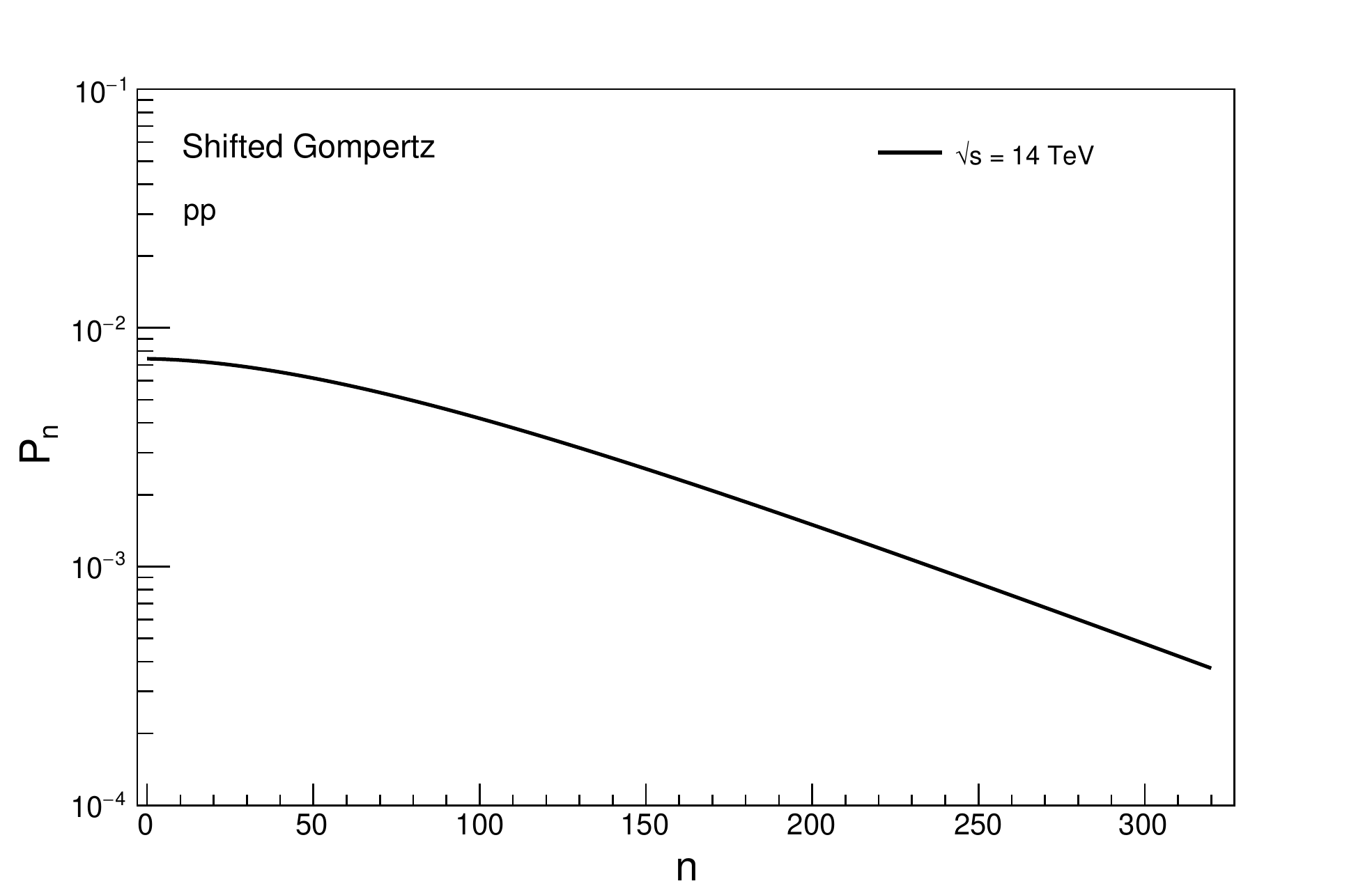}
\caption{Probability distribution $pp$ collisions predicted from shifted Gompertz function.}
\end{figure}
An interesting description of universality of multiplicity in $e^{+}e^{-}$ and $p+p(\overline{p})$ has been discussed by Grosse-Oetringhaus et al \cite{Diso}.~It is shown that although the multiplicity distributions differ between $p+p(\overline{p})$ and $e^{+}e^{-}$ collisions, their average multiplicities as a function of $\sqrt{s}$ show similar trends that can be unified using the concepts of effective energy  and inelasticity.~It is also shown that the Fermi-Landau form $<n> \sim$ $s^{1/4}$ fails to describe the $pp$ multiplicity data.~But the data is well described by $<n>=A+Blns+Cln^{2}s$.~The universality appears to be valid at least up to Tevatron energies.~The multiplicities in $e^{+}e^{-}$ and $p+p(\overline{p})$ collisions become strikingly similar when the effective energy $E_{eff}$ in $p+p(\overline{p})$ collisions, available for particle production is used. 
\begin{equation}
E_{eff} = \sqrt{s}-(E_{lead},1 + E_{lead},2), \hspace{0.6cm} \langle E_{eff} \rangle = \sqrt{s}-2\langle E_{leading}\rangle
\end{equation}
where $E_{lead}$ is the energy of the leading particle and the inelasticity $K$ is defined as $K= E_{eff}/\sqrt{s}$.~ $K$ is estimated in $p+p(\overline{p})$ collisions by comparing $p+p(\overline{p})$ with $e^{+}e^{-}$ collisions.~Given a parameterization $f_{ee}(\sqrt{s})$ of the $\sqrt{s}$ dependence of the charged multiplicity $<n>$ in $e^{+}e^{-}$ collisions, one can fit the $p+p(\overline{p})$ data with
\begin{equation}
	f_{pp}(\sqrt{s}) = f_{ee}(K.\sqrt{s}) + n_{0} 
\end{equation}
The parameter $n_{0}$ corresponds to the contribution from the two leading protons to the total multiplicity and is expected to be close to $n_{0}$=2.~One can use this fit of $p+p(\overline{p})$ data to predict the multiplicities at the LHC.~As described in reference \cite{Diso}, using a fit with equation(13), Jan Fiete Grosse-Oetringhaus et al have estimated $K$ = 0.35 $\pm$ 0.01 and $n_{0}$ = 2.2 $\pm$ 0.19.~Under the assumptions that inelasticity remains constant at about 0.35 at LHC energies and that the extrapolation of the $e^{+}e^{-}$ data with the QCD form is still reliable, authors fit the $p+p(\overline{p})$ data to predict the multiplicities at the LHC.~They find $<n>$=88.9.~We use these values of inelasticity and average multiplicity to build the multiplicity distribution at $\sqrt{s}$=14~TeV using shifted Gompertz function.

Figure~12 shows the multiplicity distribution predicted from shifted Gompertz PDF for $pp$ collisions at $\sqrt{s}$~=~14~TeV at LHC.~The mean value of the multiplicity is predicted to be $<n> \approx $89.2.  
~It is observed that in general, at all energies for different types of collisions, the multiplicity distributions can be described by shifted Gompertz function.~However the LHC data at 7000 GeV in the lower rapidity windows are an exception, whereby the fits are  statistically excluded with $CL < 0.1 \%$.~At all energies, both the scale parameter $b$ and the shape parameter $\eta$ decrease with the collision energy in the center of momentum.~In the rapidity windows, $b$ decreases with the increase in the rapidity.~The shape parameter $\eta$ increases with rapidity as it determines the width of the distribution.

The fact that multiplicity distributions at higher energies show a shoulder structure is well established.~In order to improve upon shifted Gompertz fits to the data, the multiplicity distribution is reproduced by a weighted superposition of two shifted Gompertz distributions corresponding to the soft component and the semi-hard component.~It is observed that this modified shifted Gompertz distribution improves the fits excellently and the $\chi^{2}$ values diminish by several orders.~However, distributions fail for 91 GeV $e^{+}e^{-}$ data.

~The data for $pp$ collisions at 7~TeV fail for shifted Gompertz distribution in three rapidity windows.~But the modified shifted Gompertz distribution shows a very good agreement with the data for all rapidities, as shown in table III.~For each of the rapidity bins, $\chi^{2}$/dof values are reduced manifold with $CL > 0.1 \%$. 

Using the shifted Gompertz function and the analysis, the multiplicity distributions at future collider and the mean multiplicity predicted for 500~GeV $e^{+}e^{-}$, agrees very well with the predictions from NLO QCD prediction and also with the Fermi-Landau model of particle production.

\section{Conclusion}

The aim of this paper is to propose the use of a new statistical distribution for studying the multiplicity distributions in high energy collisions; shifted Gompertz distribution function often used in model of adoption of innovations, describes the multiplicity data extremely well.~A detailed analysis of data from $e^{+}e^{-}$, $\overline{p}p$ and $pp$ collisions at high energies in terms of shifted Gompertz distribution shows that, in general the distribution fits the data very well at most of the energies and in various rapidity intervals with the exception of a very few.~Very similar to the Weibull distribution, which recently has been extensively used, it determines two non-negative parameters measuring the scale and shape of the distribution.~A power law dependence of the scale parameter and shape parameter on the collision energy is established for the $e^+e^-$ data.~The parametrisation as a power law is inspired by the observation that single particle energy distribution obeys a power law behaviour.

The occurrence of a shoulder structure in the multiplicity distribution (MD) of charged particles at high energy is well established.~This affects the shape of the distribution fit.~To improve upon the fits to the data, a weighted superposition of the distributions using shifted Gompertz function for the soft events (events with mini-jets) and the semi-hard events (events without mini-jets) is done.~The concept of superposition originates from purely phenomenological and very simple considerations.~The two fragments of the distribution suggest the presence of the substructure.~The two-component shifted Gompertz distribution fits the data from different types of collisions at different energies, extremely well.~Describing the MD in terms of soft and semi-hard components, allows one to model, under simple assumptions the new energy domain.~While predicting the multiplicity distribution using shifted Gompertz Distribution at 14~TeV, it remains interesting to determine the dependence of fraction of minijet events, $\alpha$ upon the rapidity windows compared to the events without mini-jets.~To predict the more accurate multiplicity distributions in different rapidity windows at 14~TeV, modified shifted Gompertz PDF is required, for which $\alpha$ value in each rapidity window is needed.~The analysis presented for 7~TeV data shows that the minijet fraction of events decreases with energy as well as with the increasing size of rapidity window.~This trend has also been shown in reference \cite{NBD} where the $\alpha$ fraction for full rapidity range of $pp$ collisions at 14~TeV has been estimated as 0.30.~When multiplicity distributions in full phase space at higher energies like 7 TeV, become available, the extrapolations from the lower energy domain to the highest energies can be well established, as predicted in other works also, using different approaches \cite{Cap,Fuge}.

A good agreement between the mean multiplicity and the multiplicity dependence on energy, predicted by NLO QCD and the Fermi-Landau model of particle production, with the predictions made by shifted Gompertz distribution, serves as a good test of the validity of the proposed distribution.  

The future extension of the present work shall focus on analysis of multiplicities from lower energy domains, in hadron-nucleus interactions and nucleus-nucleus interactions using shifted Gompertz distribution and to derive the additional information from the oscillatory behaviour of the counting statistics, as suggested by Wilk and W\'{l}odarczyk \cite{Wilk}.    
 
\section*{Acknowledgement}

The author M.~Kaur is thankful to Harrison Prosper of Florida State University, US for discussion on the distributions suitable to describe the collision data from the high energy accelerators. The author Ridhi Chawla is grateful to the Department of Science and Technology, Government of India for the Inspire-Fellowship grant. 

\section*{References}

\bibliography{mybibfile}

\end{document}